\documentclass[prb,twocolumn,showpacs,superscriptaddress]{revtex4}
\pdfoutput=1 %for the arxiv submission
\usepackage{epsfig}
\usepackage{amsmath,amssymb}
\usepackage{mathtools}
\usepackage{graphicx}
\usepackage{footnote}

\usepackage[normalem]{ulem}
\usepackage{color}

\usepackage[all]{xy}

\newcommand{\ceq}[1]{Eq.~(\ref{#1})}
\newcommand{\cfg}[1]{Fig.~\ref{#1}}

\begin{document}

\title{Continuous-time Quantum Monte Carlo using Worm Sampling}

\author{P.~Gunacker}
\affiliation{\small\em Institute for Solid State Physics, Vienna University of Technology, 1040 Vienna, Austria}
\author{M.~Wallerberger}
\affiliation{\small\em Institute for Solid State Physics, Vienna University of Technology, 1040 Vienna, Austria}
\author{E.~Gull}
\affiliation{\small\em Department of Physics, University of Michigan, Ann Arbor, Michigan 48109, USA}
\author{A.~Hausoel}
\affiliation{\small\em Institute for Theoretical Physics and Astrophysics, University of W\"urzburg, Am Hubland 97074 W\"urzburg, Germany}
\author{G.~Sangiovanni}
\affiliation{\small\em Institute for Theoretical Physics and Astrophysics, University of W\"urzburg, Am Hubland 97074 W\"urzburg, Germany}
\author{K.~Held}
\affiliation{\small\em Institute for Solid State Physics, Vienna University of Technology, 1040 Vienna, Austria}

\date{\small\today}
\begin{abstract}

We present a worm sampling method for calculating one- and two-particle Green's functions using continuous-time quantum Monte Carlo simulations in the hybridization expansion (CT-HYB).
Instead of measuring Green's functions by removing hybridization lines from partition function configurations, as in conventional CT-HYB, the worm algorithm  directly samples the Green's function. We show that worm sampling is necessary to obtain general two-particle Green's functions which are not of density-density type and that it improves the sampling efficiency when approaching the atomic limit. Such two-particle Green's functions are needed to compute off-diagonal elements of susceptibilities and occur in diagrammatic extensions of the dynamical mean field theory and efficient estimators for the single-particle self-energy.

\end{abstract}
\maketitle

\section{Introduction} \label{sec:Intro}
The Anderson impurity model (AIM)\cite{Anderson,Hewson} is one of the fundamental models for electronic correlations. The model was originally developed to describe the physics of magnetic impurities in solids, but nowadays  also serves as a model for quantum dots,\cite{Glazman,Ng,Kouwenhoven0} adatoms on surfaces\cite{Madhavan,Li} and appears as an auxiliary model in the context of dynamical mean field theory (DMFT).\cite{Metzner,Georges,Kotliar_dmft,Held} Continuous-time quantum Monte Carlo (CT-QMC) algorithms\cite{Rubstov_ct_int,Werner_qmc,Werner,Gull_aux,Gull} are state of the art for the numerical solution of the AIM. They are based on a stochastic sampling of an imaginary time partition function expansion.\cite{Prokofev_ct_orig,Prokofev_ct}

The methods are formally numerically exact and, in contrast to other impurity solvers,\cite{Hirsch,Sakai,Bulla,Karski,Ganahl,Caffarel,Zgid} can treat impurities with many degrees 
of freedom, general interactions, and continuous bath dispersions. The most widely known representatives are formulated as an expansion of the partition function either in terms of the interaction (CT-INT and CT-AUX)\cite{Rubstov_ct_int,Gull_aux} or in terms of the impurity-bath hybridization (CT-HYB),\cite{Werner_qmc,Werner} with the resulting series sampled stochastically.

A variant of continuous-time algorithms, usually referred to as the worm algorithm, expands both the partition function and the Green's function. This results in the configuration space sampled by Monte Carlo to be enlarged (see \cfg{figure:configuration_space}). This concept has been pioneered for diagrammatic Monte Carlo solvers for bosonic Green's functions\cite{Prokofev_ct,Prokofev_hfye} and adapted for fermionic one-particle Green's functions for the CT-INT algorithm.\cite{Burovski}

\begin{figure}
\centering
\includegraphics[scale=0.30]{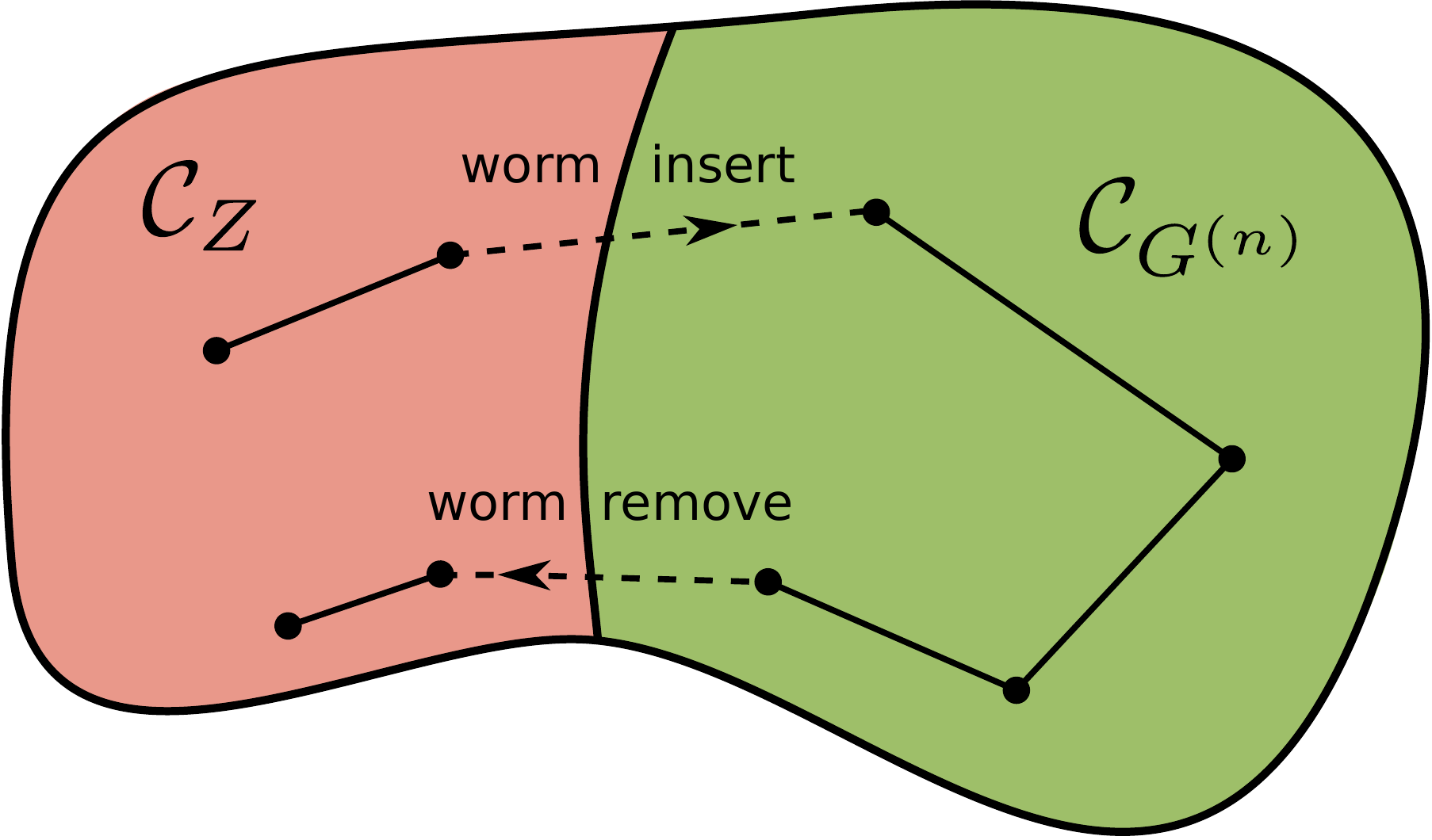}
\caption{Illustrating the concept of worm sampling. The configuration space of the partition function $\mathcal{C}_Z$ is enlarged by the configuration space of the $n$-particle Green's function $\mathcal{C}_{G^{(n)}}$. A random walk in the combined configuration space is shown, where dashed lines represent the transition moves between the two configuration spaces and solid lines the moves within one space.}
\label{figure:configuration_space}
\end{figure}

In this paper, we introduce a generalization of the worm algorithm for the (multi-orbital) hybridization expansion.\cite{Werner} While worm sampling is not restricted to any specific quantity, we show the application to fermionic two-particle Green's functions which are necessary to compute response functions and which appear in formulations of non-local extensions of the DMFT such as the dynamical vertex approximation,\cite{Toschi} the dual fermion approach,\cite{Rubtsov} the one-particle irreducible approach\cite{Rohringer_1PI} and the DMFT to functional renormalization group.\cite{Taranto}  They also appear in the measurement of single-particle self-energies using the `improved estimator'\cite{Hafermann} technique that has been shown to yield high precision estimates for the high-frequency behavior of Green's functions.

In Section \ref{sec:Mot} we motivate of our work by showing that conventional CT-HYB partition function sampling fails due to ergodicity problems when approaching the atomic limit and when calculating general two-particle Green's functions. 

Section \ref{sec:sampling} first gives a short overview of worm sampling and then generalizes CT-HYB to the Green's function space, introducing the Monte Carlo update procedure of our CT-HYB worm method. Section \ref{sec:measurement} introduces the measurement procedure. Section \ref{sec:atomiclimit} presents the results for large interactions and the atomic limit, where analytical solutions are available. Section \ref{sec:twoparticle} focuses on results for the two-particle Green's function of the two-orbital model, further validating the worm sampling algorithm by exploiting the SU(2) symmetry of the magnetic (spin) susceptibility. Section \ref{sec:conclusion} provides a brief summary.

\section{Motivation} \label{sec:Mot}
We start with a brief motivation for measuring the $n$-particle Green's functions $G^{(n)}$ with worm sampling. The Hamiltonian considered here is that of the multi-orbital AIM, which in its most general form reads:
\begin{multline}
\label{eq:anderson}
H_{\mathrm{AIM}} = \underbrace{\vphantom{\sum_{k}} \frac{1}{2} \sum_{\alpha \beta \gamma \delta} U_{\alpha \beta \gamma \delta} d^\dagger_{\alpha} d^\dagger_{\beta} d_{\delta} d_{\gamma} + \sum_{\alpha} \tilde{\varepsilon}_{\alpha} d_{\alpha}^\dagger d_{\alpha} }_{H_{\mathrm{loc}}} + \\
+ \underbrace{\sum_{k\alpha} \varepsilon_{{k}\alpha} c^\dagger_{{k}\alpha} c_{{k}\alpha}}_{H_{\mathrm{bath}}} 
+ \underbrace{\sum_{{k}\alpha\beta} \left[ V_{{k}}^{\alpha\beta} c^\dagger_{{k}\alpha} d_{\beta} + ( V_{{k}}^{\beta \alpha} )^* d^\dagger_\alpha c_{{k} \beta}\right]}_{H_{\mathrm{hyb}}}
\end{multline}
Here, $d_{\alpha}^\dagger$ ($d_{\alpha}^{\phantom{\dagger}}$) denotes the creation (annihilation) operator of an electron with spin-orbit flavor $\alpha$ on the impurity and $c_{{k\alpha}}^\dagger$ ($c_{{k\alpha}}^{\phantom{\dagger}}$) denotes the creation (annihilation) operator of an electron of momentum $k$ in the non-interacting bath that belongs to the impurity flavor $\alpha$. The impurity problem is characterized by the one-particle levels $\tilde{\varepsilon}_\alpha$, 
the interaction matrix $U_{\alpha \beta \gamma \delta}$, the bath dispersion $\varepsilon_{k\alpha}$, and the hybridization strengths $V_{k}^{\alpha\beta}$. In CT-HYB, given some inverse temperature $\beta$, the partition function $Z=\mathrm{Tr} e^{-\beta H_\mathrm{AIM}}$ of the AIM Hamiltonian \eqref{eq:anderson} is expanded in terms the hybridization $H_{\mathrm{hyb}}$. The trace then decouples into a local part described by the local impurity Hamiltonian ($H_{\mathrm{loc}}$) and a bath part described by the conduction electron bath ($H_{\mathrm{bath}}$). With the bath partition function $Z_{\mathrm{bath}}=\mathrm{Tr}{e^{-\beta H_{\mathrm{bath}}}}$ we find (for a detailed derivation see Ref.~\onlinecite{Werner_qmc}):
%\small
%\begin{multline}
\begin{align}
Z &= Z_{\mathrm{bath}} \mkern-10mu \sum_{k \in 2 \mathbb{N}_0, \alpha_k}
\int_{\tau_{k-1}}^\beta \mkern-15mu d\tau_k \int_{\tau_{k-2}}^\beta \mkern-15mu d\tau_{k-1} \ldots \int_{\tau_1}^\beta \mkern-5mu d\tau_2 \int_0^\beta \mkern-5mu d\tau_1\ \times \nonumber \\
&\times \underbrace{\mathrm{Tr}\!\left[T_\tau  e^{-\beta H_{\mathrm{loc}}} d_{\alpha_k}\!(\tau_k) d^\dagger_{\alpha_{k-1}}\!(\tau_{k-1}) \ldots d_{\alpha_2}\!(\tau_2) d^\dagger_{\alpha_1}\!
(\tau_1) \right]}_{\equiv w_{\mathrm{loc}}(k,\tau_1,\ldots,\tau_k)} \nonumber \\
&\times \underbrace{\det \mathbf{\Delta}}_{\qquad \mathclap{ \equiv w_{\mathrm{bath}}(k,\tau_1,\ldots,\tau_k)}}
\label{eq:Z}
\end{align}
%\end{multline}
%\normalsize

Here $d_{\alpha}^\dagger(\tau)$  ($d_{\alpha}^{\phantom{\dagger}}(\tau)$) are the operators of \ceq{eq:anderson} in Heisenberg representation, whose evolution in imaginary time is given by $H_{\rm loc}$. Further, $T_{\tau}$ is the Wick's time ordering operator and $\mathbf{\Delta}$ denotes the $(k/2) \times (k/2)$ matrix of all possible hybridization lines between $\tau_1 \ldots \tau_k$, with the elements:
%\small
\begin{equation}
  \label{eq:delta}
  \Delta_{\alpha \alpha^\prime}(\tau) = \sum_{{k,\gamma}}
    \frac{ (V_{{k}}^{\alpha \gamma })^* V_{{k}}^{\alpha^\prime \gamma} }
         { \mathrm{e}^{\beta \varepsilon_{k\gamma}} +1}
  \begin{cases}
      -\mathrm{e}^{-\varepsilon_{{k\gamma}} (\tau-\beta)}  & \tau > 0\\
       \mathrm{e}^{-\varepsilon_{{k\gamma}} \tau}          & \tau < 0
  \end{cases}
\end{equation}
%\normalsize
where $\tau=\tau_i-\tau_j$.
We refer to hybridizations as diagonal if $\Delta_{\alpha \alpha^\prime}(\tau) = 0 \phantom{a} \forall \alpha \neq \alpha^\prime$. Otherwise we call them off-diagonal. In the following we will restrict ourself to diagonal hybridizations, even though in principle also off-diagonal hybridizations can be considered.

In \ceq{eq:Z} abbreviations for the local weight $w_\mathrm{loc}$ and the bath weight $w_{\mathrm{bath}}$ are introduced, which become important when defining the Monte Carlo algorithm.
In \cfg{figure:hybridization} we provide an illustration of a configuration of the partition function $Z$ for a given expansion order $k/2$. A more detailed discussion is found in Ref. [\onlinecite{Werner_qmc}].
In this work we refer to the expansion order as $k/2$ so that the number of operators in the local trace is given by $k$ (different conventions exist in literature).

\begin{figure}
\centering
\includegraphics[scale=0.5]{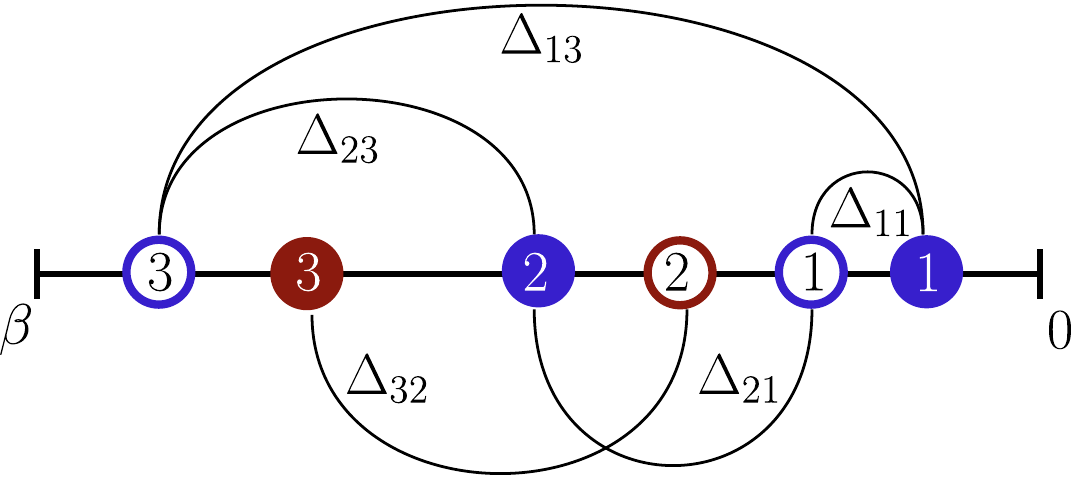}
\caption{Illustration of a configuration of the partition function $Z$ for an expansion order of $k/2=3$. 
Here we show the case of a flavor-diagonal hybridization function, which only connects operators of the same flavor to one-another. The different flavors are denoted using different colors (red, blue). When connecting creation (filled shapes) and annihilation (empty shapes) operators by hybridization lines, we number all creation operators from 1 to $k/2$ and all annihilation operators from 1 to $k/2$. }
\label{figure:hybridization}
\end{figure}

When measuring the one-particle Green's function $G^{(1)}(\tau)$ in conventional CT-HYB, sampling takes place in partition function space $\mathcal{C}_Z$, i.e., the orders $k/2$ and $\tau_i$ in \ceq{eq:Z} are sampled.  One starts from the functional identity:
\begin{equation}
  \label{eq:z_func}
  G^{(1)}_{\alpha\alpha^\prime}(\tau) = -\frac 1Z \frac{\delta Z}{\delta \Delta_{\alpha^\prime\alpha}(-\tau)}.
\end{equation}
The conventional estimator is obtained from \ceq{eq:Z} by replacing the functional derivative in \ceq{eq:z_func} with the partial derivative and using the chain rule, which generates local operators by detaching their hybridization lines (a detailed derivation can be found in Appendix C of Ref.~\onlinecite{Boehnke}):
\begin{align}
  \label{eq:est_part}
  G^{(1)}_{\alpha\alpha^\prime}(\tau) &=
   -\frac{1}{\beta} \Big\langle \sum_{nm}^{k/2} \frac{\det \mathbf{\Delta}^{(nm)}}{\det \mathbf{\Delta}} \times \nonumber \\
    &\qquad\times\ \text{sgn}\cdot \delta(\tau,\tau_m-\tau_n) \delta_{\alpha \alpha_m} \delta_{\alpha^\prime \alpha_n}\Big\rangle_{\mathrm{MC}},
\end{align} 
%In the course of this sampling one then removes all hybridization lines from a single %operator pair to measure $G^{(1)}(\tau)$. This yields the following expression for the %Green's function:\cite{Werner_qmc,Gull}
%\small
%\begin{multline}
% \begin{align}
% \label{eq:gf}
% G^{(1)}_{\alpha \alpha^\prime}(\tau) &= - \frac{Z_{\mathrm{bath}}}{Z} \mkern-20mu \sum_{k \in 2  \mathbb{N}_0, \alpha_k}
% \int_{\tau_{k-1}}^\beta \mkern-30mu d\tau_k \ldots \int_0^\beta \mkern-15mu d\tau_1 \times \det \mathbf{\Delta}^{(ij)} \mathrm{Tr} \left[ T_\tau \right.\\
% &\left. e^{-\beta H_{\mathrm{loc}}} d_{\alpha_i}(\tau) d^\dagger_{\alpha_{j}}(0) d_{\alpha_k}(\tau_k) d^\dagger_{\alpha_{k-1}}(\tau_{k-1})
%       \ldots d_{\alpha_2}(\tau_2) d^\dagger_{\alpha_1}(\tau_1) \right]
% \end{align}
%\end{multline}
%\normalsize
where $\mathbf{\Delta}^{(nm)}$ is the $(k/2) \times (k/2)$ hybridization matrix $\mathbf{\Delta}$  with the $n$-th row and $m$-th column removed, corresponding to the removal of hybridization lines;
% Let us formally rewrite \ceq{eq:gf} as a Monte Carlo estimator:
% \small
% \begin{equation}
% \label{eq:est_part}
% G^{(1)}_{\alpha_i \alpha_j}(\tau)=\frac{1}{\beta} \langle \sum_{nm}^{k/2} \frac{\det \mathbf{\Delta}^{(nm)}}{\det \mathbf{\Delta}} \text{sgn} \delta(\tau,\tau_i-\tau_j) \delta_{\alpha_i \alpha_m} \delta_{\alpha_j \alpha_n}\rangle_{\mathrm{MC}},
% \end{equation} 
% \normalsize
$\delta(\tau,\tau_m-\tau_n)$ specifies the imaginary-time bin of the measurement
and $\langle ... \rangle_{\mathrm{MC}}$ refers to the Monte Carlo expectation value of the $\tau$ integrals and $k$ sums of \ceq{eq:Z} including the weighting factor $e^{-\beta H_{\mathrm{loc}}}$; ``sgn'' denotes the sign imposed by the Wick time ordering.
In the following we will denote the estimate \ceq{eq:est_part}, suppressing the indices $\alpha,\alpha^\prime$, as $G^{(1)}_{\mathcal{C}_Z}(\tau)$.

Computing the Green's function by evaluating the quotient of the hybridization matrix reveals a first shortcoming of this approach: the estimator in \ceq{eq:est_part} fails if the hybridization between the impurity and the bath becomes very weak. We will later see that the estimator of worm sampling instead does not depend on the determinant ratio of the hybridization matrix $\mathbf{\Delta}$, such that sampling is still possible for small or vanishing $\mathbf{\Delta}$. This suggests that worm sampling does
a better job for systems approaching the atomic limit. We point out that some methods exist in order to improve the estimator in \ceq{eq:est_part}. A recent approach
is the so called remove-shift measurement (or sliding measurement), which has been implemented for density-density codes.\cite{Augustinsky} While the remove-shift estimator is capable of enhancing measurements by decreasing auto-correlation times, it does still depend on operator pairs which are connected to the bath over their hybridization. As such, this approach does not cure the problem encountered for weakly hybridizing systems.

\ceq{eq:est_part} is restricted to diagrams produced by partition function sampling and does not generate off-diagonal Green's function contributions for diagonal hybridization matrices $\Delta_{\alpha \alpha}$. However, for non-density-density interactions, such terms are indeed present in the two-particle Green's function $G^{(2)}$. One can immediately see this for the $\mathrm{SO}(n)\otimes \mathrm{SU}(2)$-conserving Slater-Kanamori interaction: here, the spin susceptibility is invariant under spatial rotations, such that, e.\,g., $\langle S_z (\tau) S_z(0) \rangle = \langle S_x (\tau) S_x(0) \rangle$.  The spin susceptibility in $z$-direction relates to flavor-diagonal terms of $G^{(2)}$:
%\begin{widetext}
\small
\begin{multline}
\label{eq:sz_sz}
\langle S_z^i(\tau) S_z^j(0) \rangle = \frac{1}{4} \langle ( n^i_\uparrow (\tau) - n^i_\downarrow (\tau) ) ( n^j_\uparrow (0) - n^j_\downarrow (0) ) \rangle = \\
\frac{1}{4} \langle c_{i \uparrow}^\dagger(\tau) c_{i \uparrow}(\tau) c_{j \uparrow}^\dagger(0) c_{j \uparrow}(0) - c_{i \uparrow}^\dagger(\tau) c_{i \uparrow}(\tau) c_{j \downarrow}^\dagger(0) c_{j \downarrow}(0) - \\
c_{i \downarrow}^\dagger(\tau) c_{i \downarrow}(\tau) c_{j \uparrow}^\dagger(0) c_{j \uparrow}(0) + c_{i \downarrow}^\dagger(\tau) c_{i \downarrow}(\tau) c_{j \downarrow}^\dagger(0) c_{j \downarrow}(0) \rangle.
\end{multline}
\normalsize
All terms can be obtained in conventional CT-HYB by removing one hybridization line for orbital $i$ and one for orbital $j$, analogous to \ceq{eq:est_part}.
The spin susceptibility in $x$-direction on the other hand manifests itself as spin flip terms in $G^{(2)}$, which are off-diagonal:
\small
\begin{multline}
\label{eq:sx_sx}
\langle S_x^i(\tau) S_x^j(0) \rangle = \frac{1}{4} \langle ( S^i_+ (\tau) + S^i_- (\tau) ) ( S^j_+ (0) + S^j_- (0) ) \rangle = \\
\frac{1}{4} \langle c_{i \uparrow}^\dagger(\tau) c_{i \downarrow}(\tau) c_{j \uparrow}^\dagger(0) c_{j \downarrow}(0) + c_{i \uparrow}^\dagger(\tau) c_{i \downarrow}(\tau) c_{j \downarrow}^\dagger(0) c_{j \uparrow}(0) + \\
c_{i \downarrow}^\dagger(\tau) c_{i \uparrow}(\tau) c_{j \uparrow}^\dagger(0) c_{j \downarrow}(0) + c_{i \downarrow}^\dagger(\tau) c_{i \uparrow}(\tau) c_{j \downarrow}^\dagger(0) c_{j \uparrow}(0) \rangle.
\end{multline}
\normalsize
%\end{widetext}
We emphasize that the two-particle generalization of \ceq{eq:est_part} does not provide the spin-flip terms of \ceq{eq:sx_sx} but only density-density-like terms as in \ceq{eq:sz_sz}.
One can obtain \ceq{eq:sx_sx} by a functional derivative as in \ceq{eq:z_func}, albeit with a hybridization function $\Delta_{\alpha \alpha^\prime}$ that is either off-diagonal in the orbitals or the spins.
Such terms are however not generated in the hybridization expansion \ceq{eq:Z}, at least not for an orbital- and spin-diagonal $\Delta_{\alpha \alpha}$.

A particularly important application of these off-diagonal elements is found when extracting the self-energy $\Sigma(i\omega)$ from the equation of motion; a technique that leads to precise high-frequency estimates. This method is usually referred to as improved estimators and has so far only been implemented for density-density interactions.\cite{Hafermann} For interactions of non-density-density type, off-diagonal terms of the two-particle Green's function are needed when implementing improved estimators for the self-energy and the reducible vertex. We will show how worm sampling is capable of supplying such off-diagonal terms, hereby overcoming the systematic shortcoming of traditional CT-HYB algorithms of being restricted to Green's functions generated by the type of AIM hybridization.

\section{Sampling and Ergodicity in Green's Function Space} \label{sec:sampling}
In order to solve the restrictions of the conventional Green's function estimator, \ceq{eq:est_part}, we may be tempted to turn the diagrammatic series of a local observable $\mathcal O$
\begin{align}
\langle \mathcal{O}(\tau) \rangle &= \frac {Z_\mathrm{bath}}Z \sum_{k \in 2 \mathbb{N}_0, \alpha_k}
\int_{\tau_{k-1}}^\beta \mkern-15mu d\tau_k \int_{\tau_{k-2}}^\beta \mkern-15mu d\tau_{k-1} \ldots \int_0^\beta \mkern-5mu d\tau_1\ \times \nonumber \\
 &\times \mathrm{Tr}\!\left[T_\tau  e^{-\beta H_{\mathrm{loc}}} \mathcal O(\tau) d_{\alpha_k}\!%
   (\tau_k) \ldots d_{\alpha_2}\!(\tau_2) d^\dagger_{\alpha_1}\!(\tau_1) \right] \nonumber \\
 &\times \det \mathbf{\Delta}
  \label{eq:expect_loc}
\end{align}
%\begin{equation}
%   \langle \mathcal{O}(\tau) \rangle = \frac{1}{Z} \mathrm{Tr} \left[ T_\tau e^{-\beta H} %\mathcal{O}(\tau)\right]
%\end{equation}
into a Monte Carlo estimator by inserting $\mathcal O$ into diagrams from the expansion of $Z$, \ceq{eq:Z}, and measuring the weight ratio. However, as already noted in Ref.~\onlinecite{Gull}, such an estimator for the Green's function is not ergodic (we will elaborate on this in Section~\ref{sec:ergomoves}).

Using worm sampling, we solve this issue by enlarging our configuration space
\begin{equation}
\mathcal{C} = \mathcal{C}_Z \oplus \mathcal{C}_{G^{(n)}}
\end{equation}
to contain both types of diagrams of \ceq{eq:Z} of the partition function space $\mathcal C_Z$ and \ceq{eq:expect_loc} of the $n$-particle Green's function space $\mathcal C_{G^{(n)}}$ (see \cfg{figure:configuration_space}). The sampling in $\mathcal C_{G^{(n)}}$ allows us to generate all diagrams for the Green's function, thereby circumventing the ergodicity problems  of both the estimator constructed from insertion of local operators and from removal of hybridization lines. While $\mathcal C_{G^{(n)}}$ was originally introduced as an auxiliary space to restore ergodicity and lower auto-correlation times for $\mathcal C_Z$, here the reverse can be argued: excursions to partition function space lower the auto-correlation times and provide the proper normalization for the Green's function (cf. Section~\ref{sec:measurement}).

In this work we restrict ourselves to sampling as $\mathcal{O}(\tau)$ in \ceq{eq:expect_loc} the one-particle Green's function and the two-particle Green's function in imaginary time $\tau$ defined by:
\begin{align}
\label{eq:greens_def}
&G_{\alpha_1 \alpha_2}^{(1)} (\tau_i,\tau_j) = -\langle T_\tau d_{\alpha_1}(\tau_i) d^\dagger_{\alpha_2}(\tau_j)\rangle \\ 
&G_{\alpha_1 \alpha_2 \alpha_3 \alpha_4}^{(2)} (\tau_i,\tau_j,\tau_k,\tau_l) = \notag \\
 &\hspace{2cm} \langle T_\tau d_{\alpha_1}(\tau_i) d^\dagger_{\alpha_2}(\tau_j) d_{\alpha_3}(\tau_k) d^\dagger_{\alpha_4}(\tau_l) \rangle
\label{eq:greens2_def}
\end{align}

Restricting worm sampling to the Green's functions space $\mathcal{C}_{G^{(n)}}$ has two reasons: (i) the one and two-particle Green's functions include almost all relevant information about the quantum impurity (see \ceq{eq:Z} -- \ceq{eq:est_part}). (ii) when sampling the one- and two-particle Green's function, we can compare  our results against the measurements in the partition function space $\mathcal{C}_{Z}$ (especially with regards to the normalization,  error bars and strong insulating cases). While a similar comparison in principle would be possible for the three-particle Green's function $G^{(3)}$, we do not consider this quantity because of the high computational effort involved and the less physical significance in comparison  to $G^{(1)}$ and $G^{(2)}$.

\begin{figure}
\centering
\includegraphics[scale=0.4]{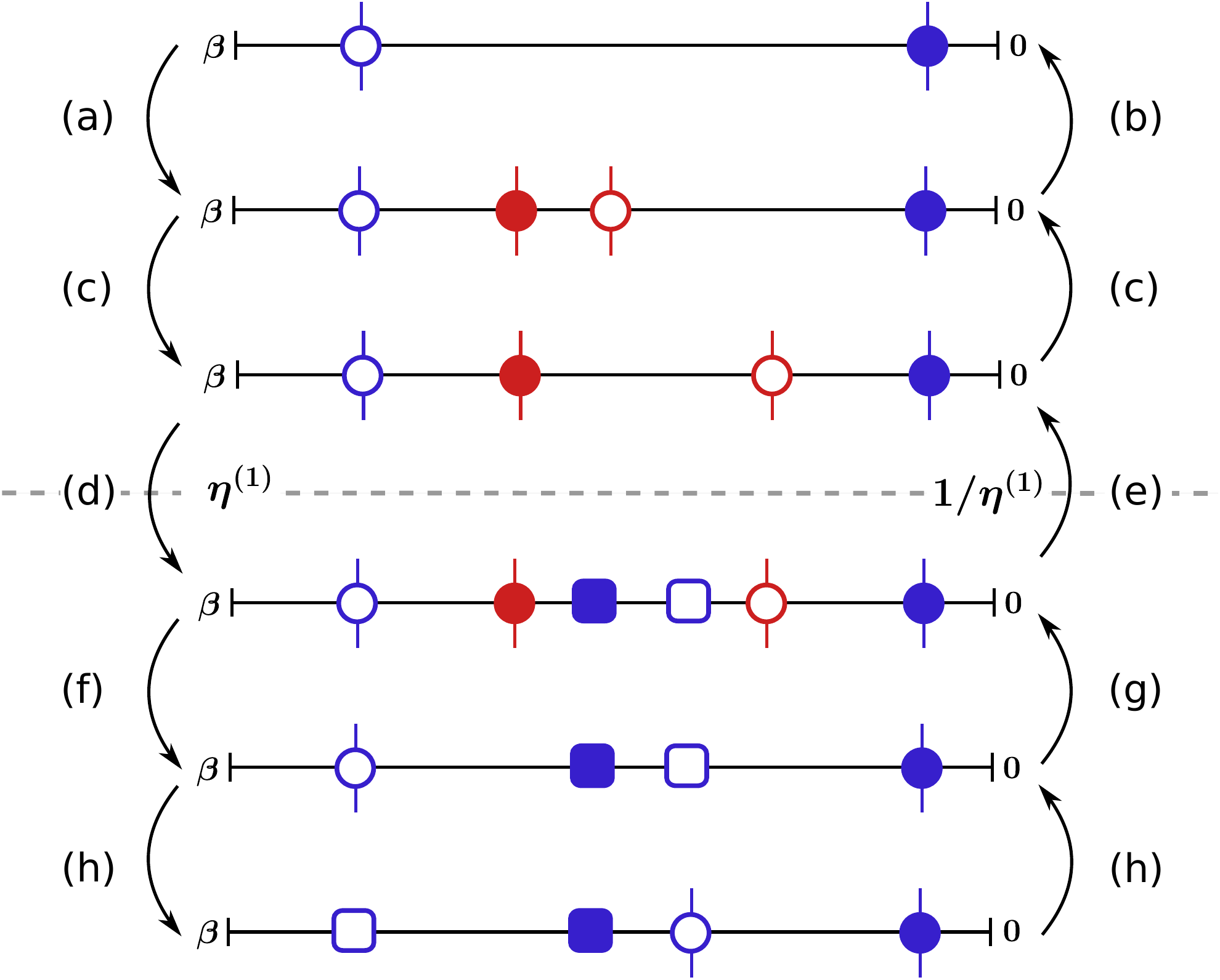}
\caption{Illustration of Monte Carlo moves in the extended configuration space of worm sampling. Circles denote operators connected by hybridization lines (indicated by vertical lines), while rectangles denote worm operators. Moves (a), (b) and (c) correspond to insertion, removal and shift of an hybridization operator pair in the configuration space $\mathcal{C}_{z}$, respectively. Labels (d) and (e) exemplify worm insertion and removal moves transitioning between the two spaces, where the parameter $\eta^{(1)}$ rescales the phase space volume of $\mathcal{C}_{G^{(1)}}$ [for details see Sec. \ref{subsec:worm_ins_rem}]. 
Labels (f) and (g) denote removal and insertion of an hybridization operator pair in $\mathcal{C}_{G^{(1)}}$ [Sec. \ref{subsec:hyb_g_ins_rem}]; (h) labels the worm operator replacement move in $\mathcal{C}_{G^{(1)}}$ [Sec. \ref{subsec:worm_rep}].}
\label{figure:moves}
\end{figure}

% Comparing the two sampling methods in $\mathcal{C}_Z$ and $\mathcal{C}_{G^{(n)}}$ then allows us to assess the advantages and disadvantages of each method.
In \cfg{figure:moves} the Monte Carlo moves in $\mathcal{C}_Z$ and $\mathcal{C}_{G^{(n)}}$ are illustrated. We included all steps needed to be ergodic and to decrease auto-correlation lengths in both configuration steps. 
The pair insertion and removal steps in $\mathcal{C}_Z$ (\cfg{figure:moves}(a),(b)) are typical in the CT-HYB algorithm.
We further introduce the operator shift move for $\mathcal{C}_Z$ (\cfg{figure:moves}(c)), which shifts the time of a creation or annihilation operator. 

For later discussion, we set up a modified partition function $Z_{G^{(n)}}$ in configuration space $\mathcal{C}_{G^{(n)}}$ by integrating over all degrees of freedom of the Green's function $G^{(n)}$:\cite{Burovski}
\begin{multline}
\label{eq:z_g_worm_partition}
Z_{G^{(n)}} := \phantom{a} \mathclap{\displaystyle\int}\mathclap{\textstyle\sum}\;\;\; G_{\alpha_1, \ldots, \alpha_n}^{(n)} (\tau_1,\ldots \tau_n)  \\ 
= \sum_{\alpha_1,\ldots,\alpha_n} \int \mathrm{d}\tau_1 \ldots \mathrm{d}\tau_n \phantom{a} G_{\alpha_1, \ldots, \alpha_n}^{(n)} (\tau_1,\ldots \tau_n). 
\end{multline}
This is not a ``physical'' partition function in the sense that it is connected to a thermodynamic potential, but it simply represents a phase space volume in Green's function space.
We will now discuss all the steps mentioned in \cfg{figure:moves} in full detail.

\subsection{Worm Insertion and Removal Steps} \label{subsec:worm_ins_rem}
The worm insertion and removal steps are transition steps
between the two configuration spaces, depicted in \cfg{figure:moves}~(d),(e). In order to sample in $\mathcal{C}_Z$ and $\mathcal{C}_{G^{(n)}}$, jumping between the two spaces is needed. 
%We give a more detailed discussion why
%it is necessary to sample in both spaces in the next paragraph.
In general, the configuration spaces $\mathcal{C}_Z$ and $\mathcal{C}_{G^{(n)}}$ have very different 
phase space volumes. This difference is balanced out by introducing a weighting factor $\eta^{(n)}$ so that the total partition function reads
\begin{equation}
\label{eq:total_config_space}
W = Z + \eta^{(n)} Z_{G^{(n)}}.
\end{equation}
For now it was not formalized how $\eta^{(n)}$ scales with the number of orbitals, temperature and interaction strength. It is best to choose $\eta^{(n)}$ so that the simulation spends
an equal amount of steps in $\mathcal{C}_Z$ and $\mathcal{C}_{G^{(n)}}$. We revisit this fact when discussing the normalization of the worm result in the
following section. 

It is important to mention that the only difference between worm operators and hybridization operators is the missing of hybridization lines. This has some implications for our Metropolis
acceptance rates. The proposal rate of inserting a worm is given by the same expression as the proposal rate of inserting $n$ hybridization operator pairs, i.e.,\cite{Werner_qmc}
\begin{equation}
f(\mathcal{C}_Z \rightarrow \mathcal{C}_{G^{(n)}}) = \frac{d \tau^{2n}}{\beta^{2n}}.
\end{equation}

Adding worm pairs results in the expansion order $k/2$ of the local trace being increased by $n$, whereas the expansion order in the determinant is kept constant. This adds an
ambiguity to the expansion order which needs to be kept in mind.
The weight of a configuration in $\mathcal{C}_{G^{(n)}}$ modified by $\eta^{(n)}$ is then:
\begin{multline}
p(\mathcal{C}_{G^{(n)}},\tau_1,\ldots,\tau_k;\tau_{i_1},...,\tau_{i_{2n}}) =  \\ 
\eta^{(n)} \cdot w_{\mathrm{loc}}(k+2n,\tau_1,\ldots,\tau_k;\tau_{i_1},...,\tau_{i_{2n}}) \\
w_{\mathrm{bath}}(k,\tau_1,\ldots,\tau_k) d\tau_1 \ldots d\tau_k.
\end{multline}

We point out that combining the proposal probability and the configuration of the weight, the $2n$ infinitesimals $d\tau_{i_1} \ldots d\tau_{i_{2n}}$ do not cancel as they would have in partition function sampling. 
This is due to the extra local degrees of freedom introduced by the worm and is integrated over in the computation of $Z_{G^{(n)}}$\eqref{eq:z_g_worm_partition}.
The proposal probability for removing the worm is simply:
\begin{equation}
f(\mathcal{C}_{G^{(n)}} \rightarrow \mathcal{C}_Z) = 1.
\end{equation}
Note, since there is only one worm in the trace at a given time, we always propose to remove exactly this worm. The Metropolis acceptance rate of a worm insertion is hence:
\begin{multline}
\label{eq:acceptance_worm_ins_1p}
a(\mathcal{C}_Z \rightarrow \mathcal{C}_{G^{(n)}}) = \\
\text{min}\!\left[1,
\eta^{(n)} \frac{\left| w_{\mathrm{loc}}(k+2n,\tau_1,\ldots,\tau_k;\tau_{i_1},\ldots,\tau_{i_{2n}}) \right|}
{\left| w_{\mathrm{loc}}(k,\tau_1,\ldots,\tau_k) \right|} 
\beta^{2n} \right]\!.
\end{multline}
The bath weight $ w_{\mathrm{bath}}$, which includes the hybridization matrix, cancels out due to the fact that the bath remains unchanged.

The inverse gives the acceptance probability of a worm removal:
\begin{multline}
\label{eq:acceptance_worm_rem_1p}
a(\mathcal{C}_{G^{(n)}} \rightarrow \mathcal{C}_Z) = \\
\text{min}\!\left[1,
\frac{1}{\eta^{(n)}} \frac{\left| w_{\mathrm{loc}}(k,\tau_1,\ldots,\tau_k) \right|}{\left| w_{\mathrm{loc}}(k+2n,\tau_1,\ldots,\tau_k;\tau_{i_1},\ldots,\tau_{i_{2n}}) \right|} \frac{1}{\beta^{2n}} \right].
\end{multline}

We point out that we jump between $\mathcal{C}_Z$ and $\mathcal{C}_{G^{(1)}}$ and between $\mathcal{C}_Z$ and $\mathcal{C}_{G^{(2)}}$, but never between $\mathcal{C}_{G^{(1)}}$ and $\mathcal{C}_{G^{(2)}}$.
As mentioned in Section \ref{sec:Mot}, the two-particle Green's function for non-density-density interaction includes spin flip and pair hopping terms. The one-particle Green's function, on the other hand, is always flavor-diagonal for flavor-diagonal hybridization
functions. This way, inserting two worm pairs consecutively by attempting to jump from $\mathcal{C}_{Z}$ to $\mathcal{C}_{G^{(1)}}$ and then to $\mathcal{C}_{G^{(2)}}$ will fail to provide flavor-off-diagonal i.e. spin flip and pair hopping terms. 
A very similar observation was recently made for the conventional CT-HYB algorithm with a flavor-off-diagonal hybridization function \cite{Olivier}.

\subsection{Pair Insertion and Removal Steps in Green's Function Space} \label{subsec:hyb_g_ins_rem}
\label{sec:ergomoves}
\begin{figure}
\centering
\includegraphics[scale=0.4]{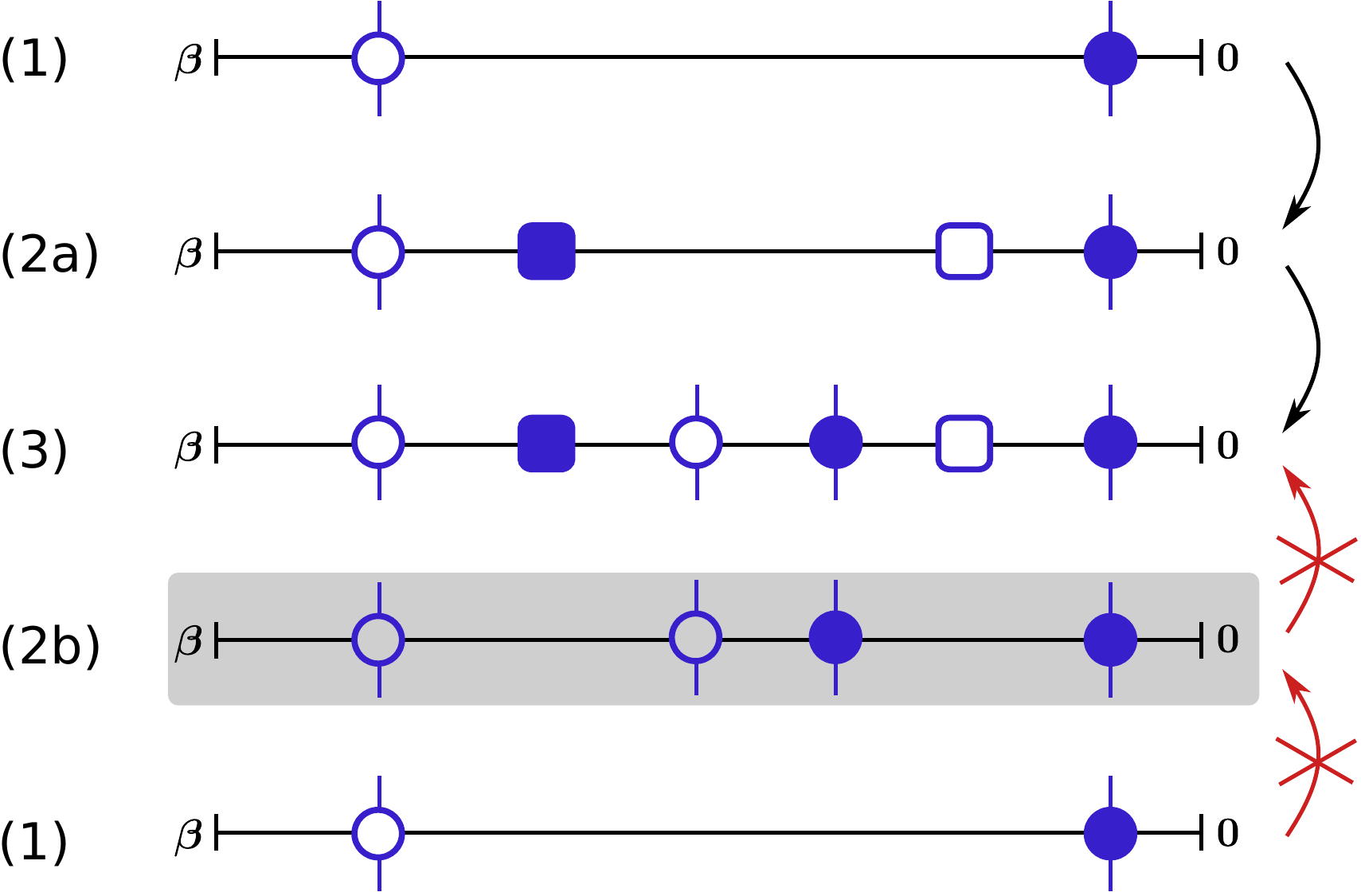}
\caption{An ``insertion estimator'', i.e. the mere insertion of local operators into a diagram from $\mathcal{C}_Z$ without sampling, is not ergodic: it fails to produce diagram (3) because (2b) violates the Pauli principle and is therefore never reached. By first transitioning to $\mathcal{C}_{G^{(1)}}$ space from (1) and then inserting a hybridization operator pair into (2a), one indeed is able to reach diagram (3).}
\label{figure:ergodicity}
\end{figure}

In order to generate all possible Green's function configurations, we need to introduce additional updates in the Green's function space $\mathcal{C}_{G^{(n)}}$.
This is a  crucial part of worm sampling: without it, the estimator is not ergodic (cf. \cfg{figure:ergodicity}).
%When sampling in the space of the operator $\mathcal{O}$, we effectively sample the partition function with the additional local operator $\mathcal{O}$ in the trace: 

This explains why we are required to sample the Green's function space $\mathcal{C}_{G^{(n)}}$ separately with operators having hybridization lines attached. To this effect, we perform insertions and removals of hybridization operator pairs also in Green's function space (\cfg{figure:moves}(f),(g)). Acceptance rates are
similar to the corresponding acceptance rates in $\mathcal{C}_Z$ space:
\small
\begin{multline}
\label{eq:acceptance_ins_g}
a(\mathcal{C}_{G^{(n)}};k + 2n \rightarrow k + 2n +2) = \\
\text{min} \left(1, \vphantom{\int} \right. \frac{\left| w_{\mathrm{loc}}(k + 2n + 2,\tau_1,\ldots, \tau_{i_1}, \ldots, \tau_{i_{2n}}, \ldots, \tau_k; \tau_i, \tau_j) \right|}{ \left|w_{\mathrm{loc}}(k+2n,\tau_1, \ldots, \tau_{i_1}, \ldots, \tau_{i_{2n}}, \ldots,  \tau_k) \right|} \times\\
\left. \frac{\left|w_{\mathrm{bath}}(k+2,\tau_1,\ldots,\tau_k;\tau_i,\tau_j) \right|}{\left| w_{\mathrm{bath}}(k,\tau_1,\ldots,\tau_k) \right|} 
\frac{\beta^2}{((k+2)/2)^2} \right),
\end{multline}
\normalsize
where the worm operators are located at times $\tau_{i_1}, \ldots, \tau_{i_{2n}}$. The Metropolis acceptance rate for a pair removal in the Green's function space 
is then just given by the inverse of \ceq{eq:acceptance_ins_g}.

We remind the reader of the fact that the local weight $w_{\mathrm{loc}}$ in \ceq{eq:acceptance_ins_g} is expressed relative to a factor $k+2n+2$, while
the bath weight $w_{\mathrm{bath}}$ is expressed relative to a factor $k+2$. The discrepancy comes from the $n$ worm operator pairs in the local trace without hybridization lines.

\subsection{Worm Replacement Step in Green's Function Space} \label{subsec:worm_rep}
While insertion and removal moves formally fulfill the condition of ergodicity, worm sampling requires a shift/replacement move in order to allow for acceptable auto-correlation lengths.
We elaborate on this requirement here. 

Let us assume a local trace filled with hybridization
operator pairs. We now attempt to insert a worm pair into this trace. It turns out that inserting a worm
pair, where the worm operators are relatively close to one another is probable, while inserting a worm pair where the worm operators are far apart is less probable.
This is because of (i) possible quantum number violations since there may be many creation and annihilation  in between the pair for long time differences, and (ii) the pair insertion might lead to an energetically disadvantageous local configuration
which is unfavorable to have for a long time.

Problem (i) is especially severe if we have a large amount of operators in the trace, which occurs at small interaction or low temperatures. Additionally, more restrictive
interaction types, such as the density-density interaction, produce more rejects due to quantum number violations of attempted worm inserts. This is why we do not observe this
auto-correlation problem at high temperatures, high interaction parameters and more general interactions such as Slater-Kanamori interactions (which may change the quantum number in the local trace).

The solution to this problem is found in shift/replacement moves. We consider, instead of a general worm shift move, 
a replacement move which exchanges one of the worm operators with an operator of the hybridization expansion, i.e., we replace it 
with one  of the same flavor connected by a hybridization
line  as illustrated in \cfg{figure:moves}(h).

This way we do not have to recalculate the local trace, as two locally indistinguishable operators switch position. Instead, we need to recalculate the determinant of the
hybridization matrix since the replacement corresponds to a shift of the worm operator and a shift of the hybridization operator. Further we do not encounter any rejects of
proposed moves due to local quantum number violations.

It turns out that worm replacement moves (or in the same way worm shift moves) are equally important for traces with very few operators because of problem (ii). This problem typically occurs if the weight $e^{-U \tau}$ of the 
worm becomes prohibitively small, i.e., in particular for 
a large interaction strength and a long $\tau$ difference such as $\frac{\beta}{2}$.
We are then effectively restricted to inserting operator pairs into the trace, which are very close to each other in imaginary time. These pairs have similar
properties as density operators and can in principle be inserted for very high insulating cases. By inserting hybridization pairs at short distances $\tau_i-\tau_j$ and then replacing
one worm operator with one hybridization operator we are able to pass this restrictions of the time evolution.
As we will show in the following, the replacement move only depends on the ratio of the determinant of the hybridization matrix.

The proposal probability of a worm replacement step is given by:
\begin{equation}
\label{eq:worm_rep_prop}
f'(\mathcal{C}_{G^{(n)}},k + 2n \rightarrow k + 2n) = \frac{1}{2n(k/2)}.
\end{equation}

This corresponds to selecting one creation/annihilation operator of the $2n$ worm operators at random and selecting one creation/annihilation of the same spin-orbit flavor with a hybridization line. In practice, we choose an operator from
the $k/2$ operators of the same type (annihilator/creator) and then discard flavors, which are not equivalent to the worm flavor.
The proposal probability of switching the operators back to their original position is hence also given by \ceq{eq:worm_rep_prop}.

We observe that the proposal probabilities for the replacement move cancel out and the acceptance ratio is fully determined by the ratio of weights.
Further, the local weights cancel, since a worm operator and the corresponding hybridization operator are indistinguishable within the local trace.
The Metropolis acceptance rate is hence given by:
\begin{multline}
\label{eq:acceptance_replacement}
a'(\mathcal{C}_{G^{(n)}},k + 2n \rightarrow k + 2n) = \\
{\rm min} \left(1,
\frac{\left| w_{\mathrm{bath}}(k,\tau_1,\ldots,\tau_i,\ldots,\tau_k) \right|}
{\left| w_{\mathrm{bath}}(k,\tau_1,\ldots,\tau_j,\ldots,\tau_k) \right|} \right).
\end{multline}
where $\tau_i$ refers to the initial position of the worm operator and $\tau_j$ to the initial position of the operator with the hybridization line. 
\cfg{figure:plot_replacement} shows how worm replacement moves alleviate the ergodicity problem of the worm algorithm for the situation where many operators are
found in the local trace.

\begin{figure}
\centering
\includegraphics[width=8cm]{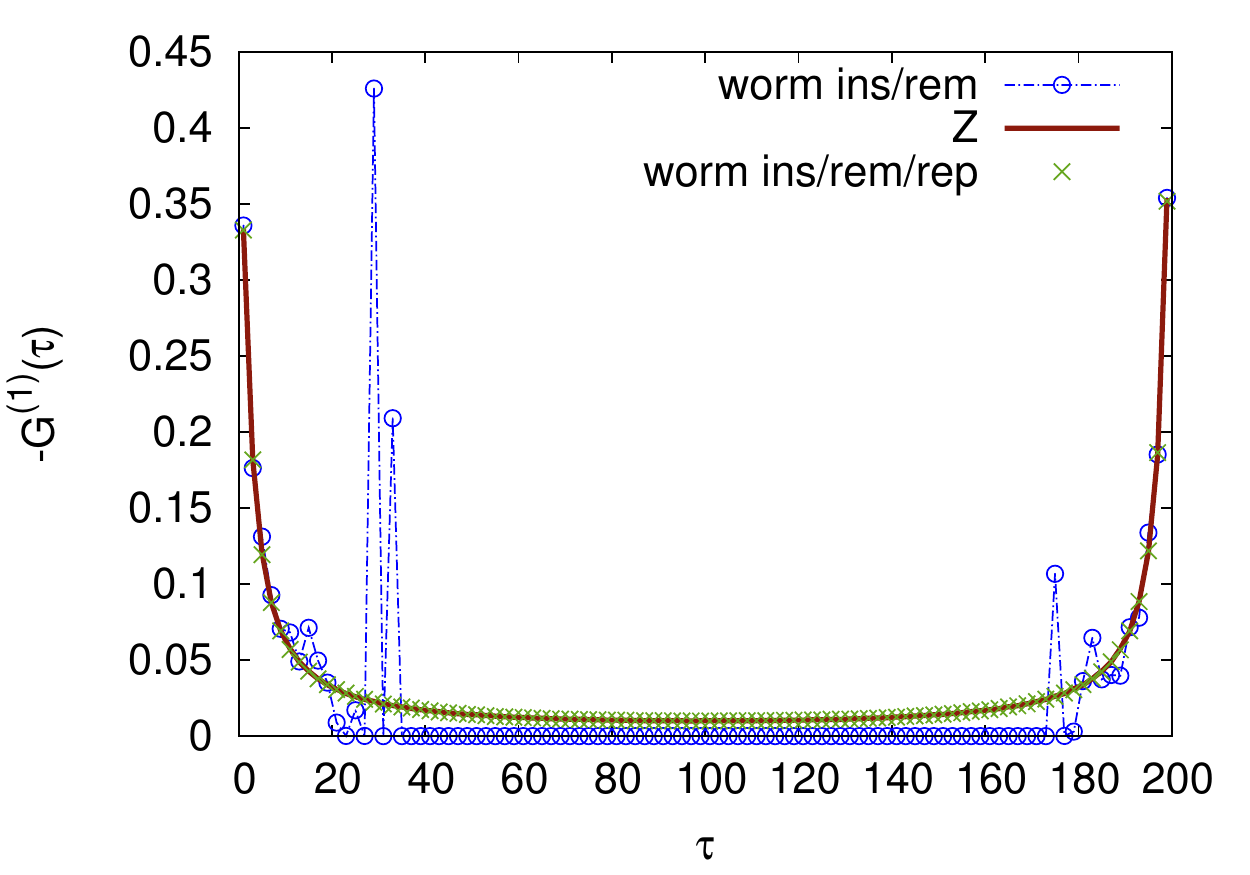}
\caption{One-particle Green's function $G^{(1)}(\tau)$ in imaginary time $\tau$, illustrating the ergodicity problem of the worm algorithm for an average expansion order of $k/2 \sim 40$. Parameters: inverse
temperature $\beta = 200/D$, Coulomb repulsion $U=0.5D$ and $\mu=0.3D$ (out of half-filling) for the the single-orbital AIM with semi-elliptic conduction electron density of states with half-bandwidth $D=1$ and $V=0.5D$. 
The balancing parameter $\eta^{(1)}$ was chosen in the interval $[0.15,0.22]$.
We observe the ergodicity problem between $\tau=25/D$ and $\tau=175/D$ (blue curve). When adding replacement moves, we are able to insert worm operators for such $\tau$'s around $\beta/2$ (green triangles) and hence obtain much better results.
We have additionally supplied $G^{(1)}(\tau)$ for the measurement in partition function space (red curve).}
\label{figure:plot_replacement}
\end{figure}

We would like to use the opportunity to point out the difference between a worm
replacement and a worm shift move. The acceptance rate of the worm replacement move depends on a determinant ratio of two matrices of dimension $(k/2) \times (k/2)$, where $k$ here refers to the number 
of operators with hybridization lines connected. In that sense it is very comparable to the determinant ratio of two matrices of dimension $(k/2 - 1) \times (k/2 - 1)$ and $(k/2) \times (k/2)$ in \ceq{eq:est_part} when changing  
the order between $k/2$  and $(k/2 - 1)$
in  partition function sampling. The acceptance rate of a worm shift move, on the other hand, only depends on the ratio
of the local traces. While for the worm replacement move we are able to pass the restrictions of the local time evolution, for the worm shift move we are able to pass
the restrictions of the hybridization function. When calculating strong insulating cases we profit the most if we consider both moves.
 
\section{Worm Measurement}\label{sec:measurement}
We now show how the measurement of Green's function looks in $\mathcal{C}_{G^{(n)}}$. It turns out that the measurement itself is trivial and we only need to find the 
correct normalization of the Green's functions measured and the correct sign. 
For the one-particle Green's function $G^{(1)}$ a worm is defined by the operators $d (\tau_i)$ and $d^\dagger (\tau_j)$. The correct weight is intrinsically given as we sample
in the Green's function space $\mathcal{C}_{G^{(n)}}$. Thus, the estimator of the Green's function simply follows as:
\begin{equation}
\label{eq:green_tau}
G_{\mathcal{C}_G}^{(1)} (\tau) = \langle \text{sgn} \cdot \delta (\tau, \tau_i -\tau_j) \rangle_{\mathrm{MC}}.
\end{equation}
  
The Green's function in Matsubara frequencies can be calculated by substituting the ${\delta\text{-function}}$ by the Fourier transform:
\begin{equation}
\label{eq:green_iw}
G_{\mathcal{C}_G}^{(1)} (i\nu) = \langle \text{sgn} \cdot e^{i\nu(\tau_i -\tau_j)} \rangle_{\mathrm{MC}}.
\end{equation}

The measurement of the two-particle Green's function in Matsubara frequencies in the particle-hole channel is given by:
\begin{multline}
\label{eq:green2_iw}
G_{\mathcal{C}_G}^{(2)} (i\nu, i\nu',i\omega) = \\
\langle \text{sgn} \cdot e^{i\nu (\tau_i - \tau_j) } e^{i\nu' (\tau_k - \tau_l) } e^{i\omega (\tau_i -\tau_l) } \rangle_{\mathrm{MC}}.
\end{multline}

The imaginary time arguments $\tau_i,\ldots,\tau_l$ are assigned to creation and annihilation operators according to \ceq{eq:greens2_def}.

While we both employ \ceq{eq:green_tau} and \ceq{eq:green_iw} for the one-particle Green's function measurement,
the measurement of the two-particle Green's function in Matsubara frequencies, \ceq{eq:green2_iw}, is far more convenient than a binned measurement in imaginary time.
It is especially difficult to resolve jumps in the imaginary-time measurement due to fermionic sign changes in the time ordering of operators. Measuring the two-particle
Green's function in imaginary time using a binning procedure and then applying the Fourier transform gives wrong high frequency asymptotics, while the direct measurement
in Matsubara frequencies is free of errors resulting from binning. 

As with conventional sampling, we do not observe any sign-problem for worm sampling in the case of a flavor-diagonal hybridization function. However, unlike in the $G^{(n)}_{\mathcal{C}_Z}$ estimator,
the flavor indices and the imaginary time bins in the worm estimator $G^{(n)}_{\mathcal{C}_G}$ are outer indices, such that the mean sign in principle also becomes flavor and $\tau$ dependent. 

\ceq{eq:green_tau} and \ceq{eq:green_iw} are normalized to $Z_{G^{(1)}}$, \ceq{eq:green2_iw} to $Z_{G^{(2)}}$ as defined in \ceq{eq:z_g_worm_partition}, as opposed to the physically correct normalization to $Z$.
We will now discuss the normalization in more detail.

\subsection{Normalization and Auto-Correlation}
In principle we are ergodic in $\mathcal{C}_{G^{(n)}}$, when
assuming worm replacement or worm shift moves. It turns out however that we need to sample both in $\mathcal{C}_{G^{(n)}}$ and $\mathcal{C}_Z$ with about the same number of steps to fix the 
normalization $\frac{1}{Z}$ of the thermal expectation value in \ceq{eq:expect_loc}. 

When measuring the Green's functions in $\mathcal{C}_{G^{(n)}}$ we implicitly normalize with the number of steps taken in $\mathcal{C}_{G^{(n)}}$. We correct for this factor
by explicitly counting how many steps $N_G$ were taken in $\mathcal{C}_{G^{(n)}}$. We further count how many steps $N_Z$ were taken in $\mathcal{C}_Z$. This estimates the size of the
configuration space $\mathcal{C}_Z$, which then gives the correct normalization. The normalization for $G^{(n)}$ is then given by:\cite{Gull_bold}
\begin{equation}
\label{eq:normalization_wz}
G^{(n)} = \frac{1}{\eta^{(n)}} \frac{N_G}{N_Z} G_{\mathcal{C}_G}^{(n)},
\end{equation}
where $G_{\mathcal{C}_G}^{(n)}$ is measured in $\mathcal{C}_{G^{(n)}}$ and the factor $1/{\eta^{(n)}}$ is a result of rescaling $Z_{G^{(n)}}$ in \ceq{eq:total_config_space}.

Let us note that \ceq{eq:normalization_wz} is only one way of normalizing the worm measurement. In a different approach, we could  do
the entire sampling in worm space, without removing the worm operators at all. We are then required to generate worm configurations by shift moves and replacement moves. In this case, we could normalize
the result by assuming some physical knowledge of the Green's function. One possibility is to extract the normalization by assuming the correct behavior of the large-frequency asymptotics of $G^{(1)}(i\nu)$ or $G^{(2)}(i\nu,i\nu',i\omega)$.    

In order to calculate the Monte Carlo expectation value \ceq{eq:green_iw}, we still need to divide by the number of measurements $N$ taken. It is important to notice the difference
between the number of measurements $N$ and the number of steps $N_{G}$ and $N_{Z}$ taken since it is common to skip steps during two consecutive measurements to assure uncorrelated measurements.

This directly relates to the auto-correlation length of the QMC sampling. The auto-correlation length in worm space $\mathcal{C}_{G^{(n)}}$ looks very different from the auto-correlation in partition function space $\mathcal{C}_Z$.
A well-accepted estimate for the auto-correlation length in traditional CT-HYB is given by the quotient of the number of operator pairs $(k/2)$ over the acceptance rate for removal in partition function space $r_{\mathrm{rem},Z}$:\cite{Gull}
\begin{equation}
\label{eq:auto_corr_z}
N_{\mathrm{corr},Z} \approx \frac{(k/2)}{r_{\mathrm{rem},Z}}.
\end{equation}

In principle, a similar estimate holds for the Green function sampling
$\mathcal{C}_{G^{(n)}}$. However, another possibility 
to arrive at an uncorrelated worm is to  remove one worm and insert a new worm into the local trace at another location.  
If the acceptance rate for removal of a worm pair is $r_{\mathrm{rem},W}$, this gives another estimate  for the auto-correlation length in worm space:
\begin{equation}
\label{eq:auto_corr_w}
N_{\mathrm{corr},W} \approx \frac{1}{r_{\mathrm{rem},W}},
\end{equation}
which we employ in practice.

It is still necessary to modify the approximations in \ceq{eq:auto_corr_z} and \ceq{eq:auto_corr_w} by the percentage of worm steps
proposed and the percentage of hybridization operator steps proposed, since our new system has two different types of moves instead of one. 
We observe that the acceptance rate of worm inserts and worm removals is in general lower when inserting four operators at once, as  is the case for the two-particle Green's function $G^{(2)}$.
While we are able to alleviate this problem partially by adjusting $\eta^{(2)}$, the acceptance rate is still lower due to quantum number violations. 
The reduced acceptance rate directly translates to an increased auto-correlation length of the two-particle Green's function.

\section{Atomic Limit Results}\label{sec:atomiclimit}
As a first test and validation of the worm algorithm we consider the atomic limit. We distinguish two scenarios with a divergent
ratio of Coulomb repulsion to hybridization strength $U/V \rightarrow \infty$. (i)  The actual atomic limit defined as $V\rightarrow 0$ for  finite $U$, i.e., we decouple the impurity from the bath. 
 In this scenario, we are still able to choose $U$ freely. This allows us to
control the time evolution in the local trace. We observe that the Green's function estimators of partition function sampling fail completely in this case due to the absence of the hybridization function.
In the second scenario (ii), we keep $V$ fixed and increase the Coulomb repulsion $U \rightarrow \infty$. While the Green's function estimator of
partition function sampling is still capable of producing results for large $U$ due to the presence of the hybridization function, we observe systematic deviations of the error bars around $\tau=\beta/2$.

\subsection{Atomic limit $ V \rightarrow 0 $}
The one-particle Green's function $G^{(1)}$ and the two-particle Green's function $G^{(2)}$ are known analytically in the atomic limit. On the other hand, estimators of the type \ceq{eq:est_part} fail completely
since the impurity is no longer coupled to the bath. That is, measuring the Green's functions
by cutting hybridization lines in CT-HYB is no longer possible due to the absence of the hybridization function. The worm algorithm, on the other hand, is not limited by the hybridization function, as operators
are inserted locally. As a result, the worm algorithm is capable of reproducing the atomic limit.

While sampling the atomic limit with QMC algorithms is mainly of academic interest, we can use the analytic results for benchmarking.
\cfg{figure:atomic_giw} shows the Green's function in the atomic limit, i.e., for an isolated  impurity, comparing  the worm algorithm and the analytic expression. 

\begin{figure}
\centering
\includegraphics[scale=0.6]{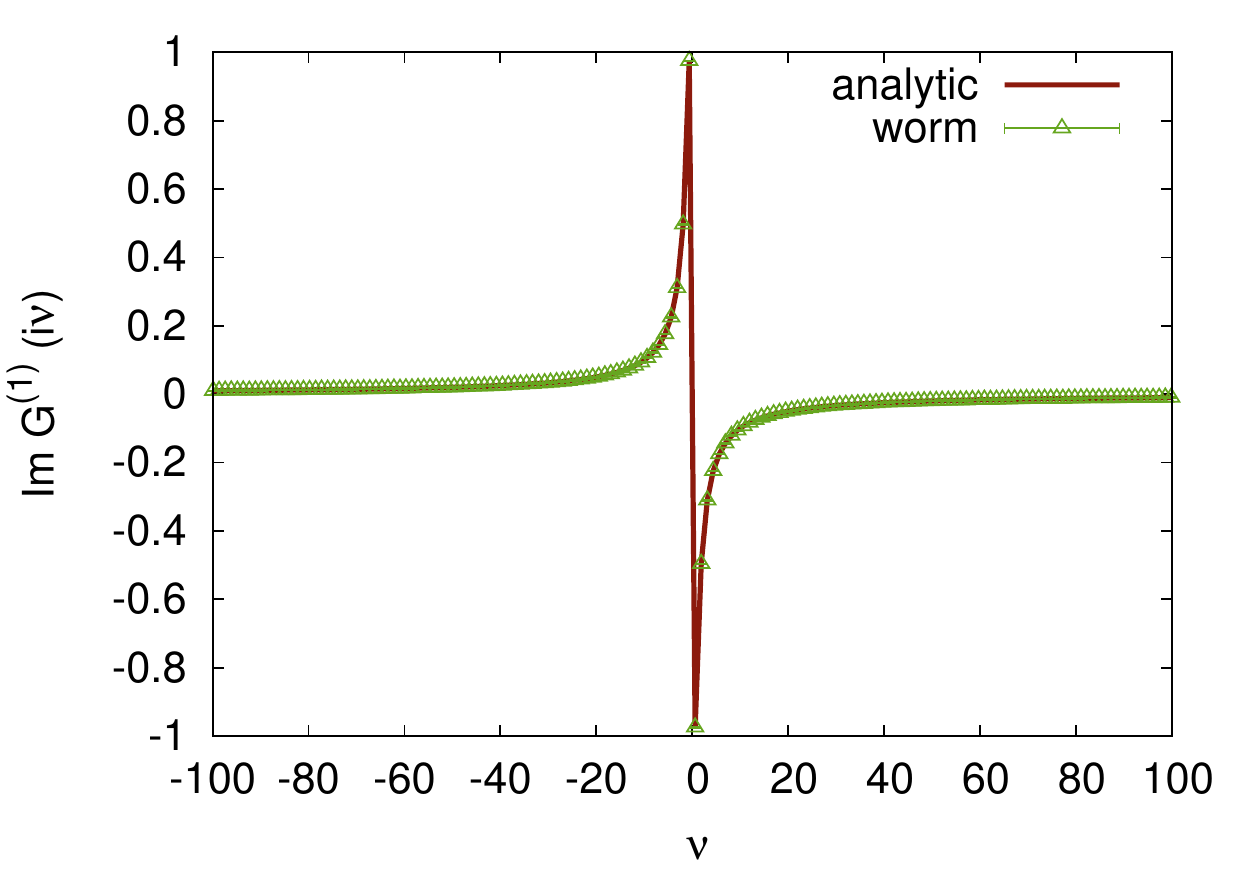}
\caption{One-particle Green's function $G^{(1)}(i \nu)$ in discrete Matsubara frequencies $i\nu$ for the atomic limit of the single-orbital AIM at  inverse temperature
$\beta = 5/D$, Coulomb repulsion $U=1.0D$ and $\mu=0.5D$ (half-filling). The balancing parameter was set to $\eta^{(1)}=0.7$. 
In the absence of any hybridization function, the worm algorithm (green triangles) is able to reproduce the analytic result (red line). Conventional CT-HYB is not possible.} 
\label{figure:atomic_giw}
\end{figure}

Let us now turn our focus towards two-particle quantities. The measurement of four worm operators in imaginary time is Fourier transformed into Matsubara frequencies using the particle-hole convention. 
The two-particle Green's function $G^{(2)}(i\nu,i\nu',i\omega)$ in the particle-hole convention is a function of two fermionic Matsubara frequencies $i\nu,i\nu'$ and one bosonic Matsubara frequency $i \omega$. 
In order to quantify results, we
analyze slices of the full two-particle Green's function by setting the second fermionic frequency to $\nu' = \pi/\beta$ and the bosonic frequency to $\omega = 0$. For comparison,
we construct the analytic atomic limit results of the two-particle Green's function from the expressions of the reducible vertex.\cite{Rohringer, Hafermann_atomic}
A more complete discussion of the general properties of two-particle quantities can be found elsewhere.\cite{Rohringer}
\cfg{figure:atomic_g4iw} shows the $G^{(2)}_{\uparrow \uparrow \downarrow \downarrow}(i \nu, \pi/\beta, 0)$ slice measured using worm sampling and compared to the  analytic result.

We conclude that the absence of the hybridization function in the atomic limit results in a complete breakdown of the one- and two-particle Green's function estimator in partition function sampling. 
In contrast worm sampling works very well and correctly reproduces the analytic result for the atomic limit.

\begin{figure}
\centering
\includegraphics[scale=0.6]{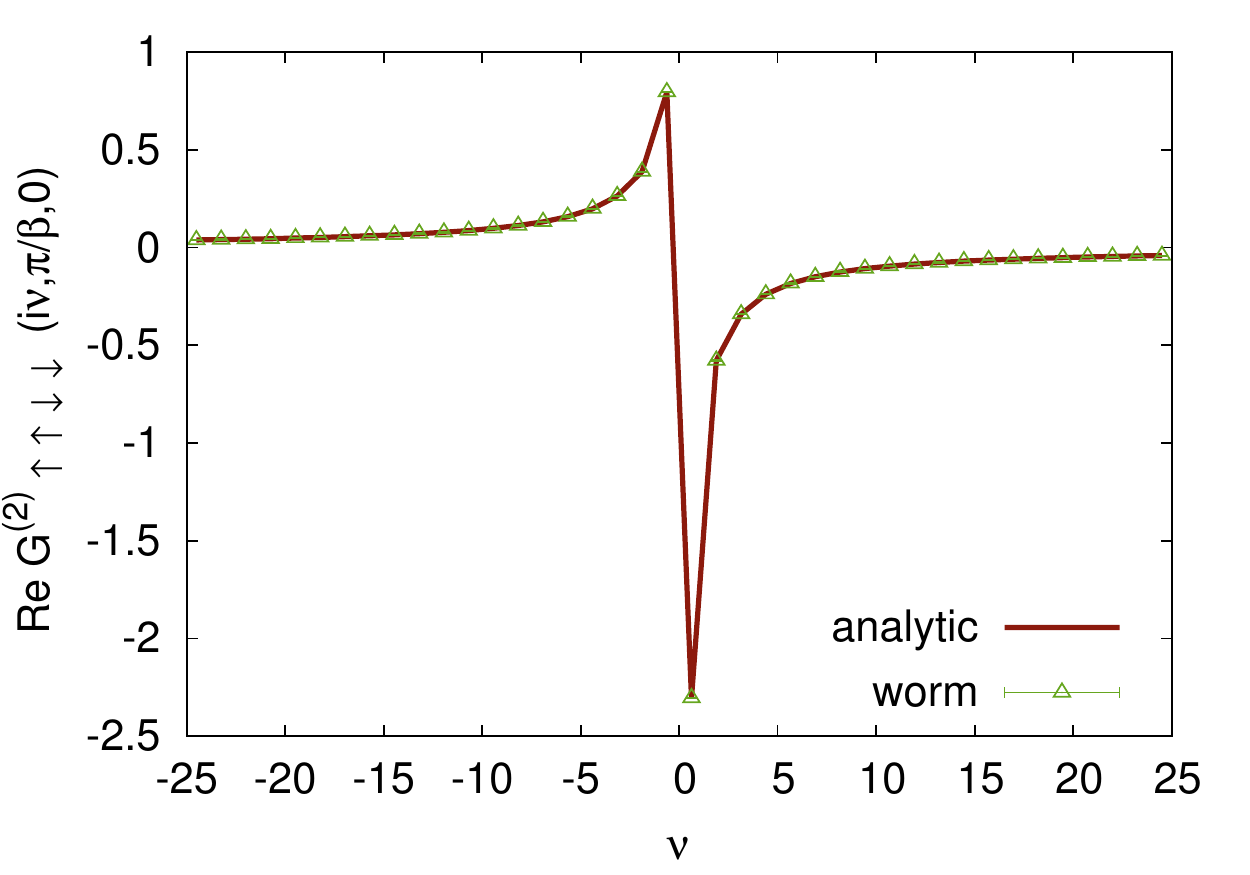}
\caption{Same as Fig.\ \ref{figure:atomic_giw} but now for the two-particle Green's function $G^{(2)}_{\uparrow \uparrow \downarrow \downarrow}(i \nu, \pi/\beta, 0)$ slice for which a  balancing parameter  $\eta^{(2)}=0.155$ has been employed.} 
\label{figure:atomic_g4iw}
\end{figure}

\subsection{Strong interaction limit $U\rightarrow \infty$}
In principle, CT-QMC algorithms are used for intermediate parameter ranges, but not the atomic limit itself. However, the strongly insulating case with high values of $U$ is of interest.
While here a hybridization function is still present for a finite bandwidth, the local time evolution suppresses most of the hopping from and onto the impurity.

\cfg{figure:strong_gtau_log} shows the one-particle Green's function $G^{(1)}(\tau)$ with error bars on a logarithmic scale. Both approaches, partition function and worm sampling, essentially agree for the Green's function. 
However, the error bars of  partition
function sampling vanish for intermediate $\tau$-values. This is clearly an artifact since the error bars should be comparable along the whole range of $\tau$-values, as it is the case in worm sampling. 
The origin for this shortcoming is that hybridization pairs for intermediate $\tau$-values are no longer inserted, but just measured by cutting hybridization lines between operators of two operator pairs.
%The origin for this shortcoming is the fact that for very large $U$ the system is most of the time in its ground state, i.e. only operator pairs of small $\Delta \tau$ are inserted or removed in partition function sampling. Thus the $\tau$ positions of this local pairs to each other are only governed by $H_{\mathrm{bath}}$. This can be seen in \cfg{figure:strong_gtau_log}, where at $\tau$ close to $0$ or $\beta$ the errorbars are good, whereas around $\tau=5$ and $\tau=45$ they become very large, because very few pairs of that $\tau$ distance were inserted. 
%No pairs of $\Delta \tau > 10$ could be inserted, thus there is no influence of $H_{\mathrm{loc}}$ on the result in between $\tau=10$ and $\tau=40$.
While the effect on the Green's function itself is still small, it  already produces wrong error bars and hence maximum entropy spectra. 
Small errors  may also propagate and get enlarged 
through DMFT iterations. 

\begin{figure}
\centering
\includegraphics[scale=0.6]{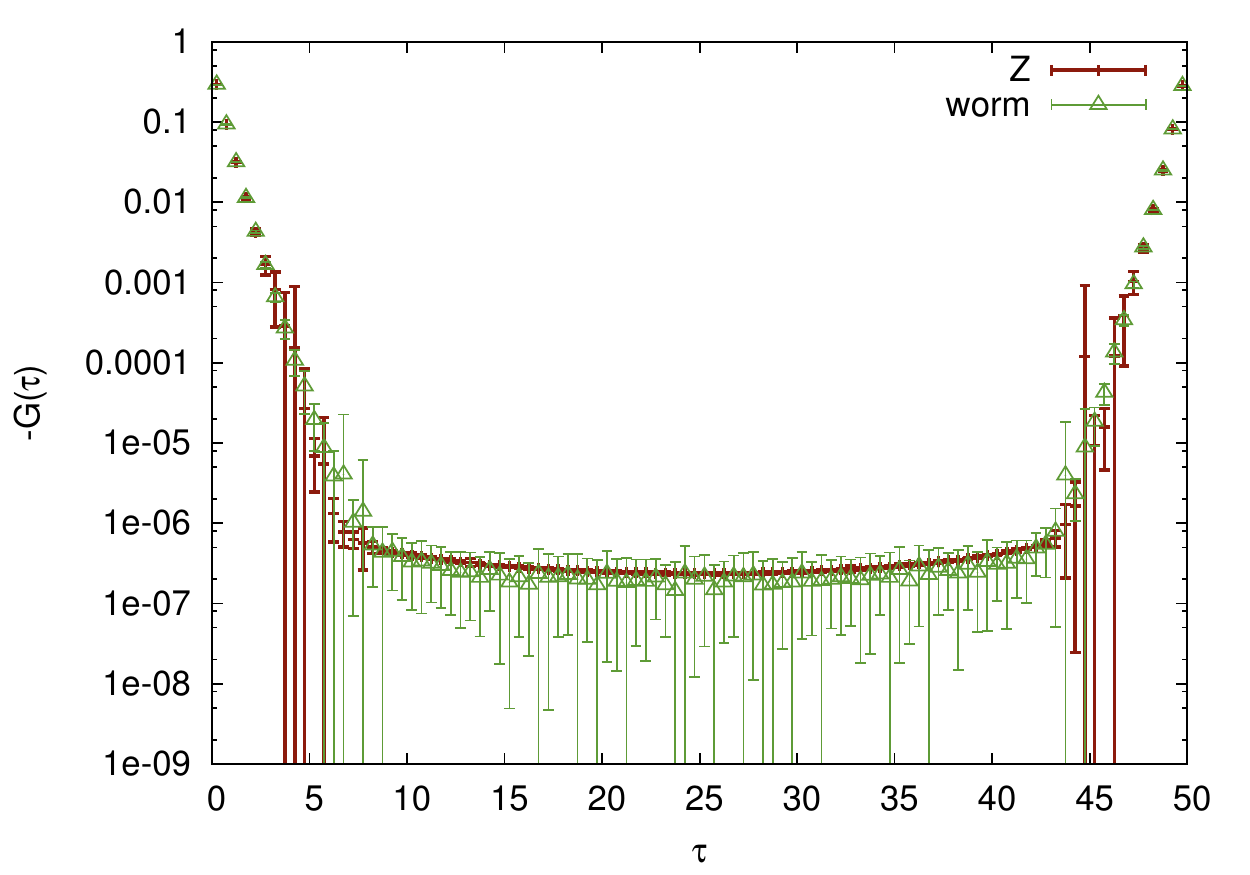}
\caption{One-particle Green's function $G^{(1)}(\tau)$ in imaginary time for the single-orbital AIM 
with semi-elliptic conduction electron density of states of half-bandwidth  $D$, $U=5.0D$ and $\mu=2.6D$ (out of half-filling). The balancing parameter was set to $\eta^{(1)}=1.4$.
While the error bars of the Green's function calculated by  partition function sampling vanish between $\tau \in [5,45]$ (red curve), the error bars of the Green's function in worm sampling have a comparable magnitude for
all values of $\tau$.} 
\label{figure:strong_gtau_log}
\end{figure}

We hence conclude that the worm algorithm not only correctly reproduces the  atomic-limit but also works properly for large $U$, including error bars. As such, the worm algorithm provides an improvement to the conventional 
CT-HYB algorithm in the strong coupling limit. It also correctly reproduces the non-interacting limit making it, in principle, numerically exact over the complete parameter range. 

\section{Two-particle Green's function}\label{sec:twoparticle}
In the previous section we have discussed how Green's function estimators in partition function sampling lead to systematic errors in the absence of a hybridization function.
This is true for any type of hybridization function. Another problem arises when dealing with spin-orbital diagonal hybridization functions. 
Such a diagonal hybridization is exact in high-symmetry cases and is a widely employed approximation in other systems, because it mitigates the sign problem and allows for speed-ups 
due to the block-diagonalization of matrices.\cite{Haule_2007,Parragh}
The CT-HYB algorithm is then only inserting operator pairs within the hybridization expansion where creation and annihilation operators have the same spin-orbit flavor.
In conventional CT-HYB partition function space sampling, Green's function estimator are measured by removing these hybridization lines. That is, one can only measure Green's functions,
which can be built from hybridization pairs with the same spin-orbit flavor.
While the one-particle Green's function in general fulfills this criteria and can be measured with such estimators (note that the flavor-off-diagonal one-particle Green's function vanishes for flavor-diagonal hybridization), 
this is not true for all components of the two-particle Green's function.

Especially the spin flip and pair hopping terms of the two-particle Green's function are not accessible in this way. This is another systematic weakness of conventional CT-HYB partition function sampling. The worm sampling algorithm, on the other hand,
does not suffer from this shortcoming. This is because  four arbitrary operators can be inserted into the trace. Their spin-orbit flavor can be chosen freely without the need to connect these via  the hybridization function.

In order to analyze the spin flip and pair hopping terms of worm sampling, we again look at the atomic limit. We choose the two-orbital AIM with semi-elliptic conduction electron density of states and  Slater-Kanamori interaction.\cite{Kanamori,Parragh}
This  local interaction includes an  intra-orbital repulsion $U$,
SU(2)-symmetric Hund's exchange and pair hopping terms  $J$,
and inter-orbital interaction  $U'=U-2J$. \cfg{figure:atomic_spinflip} and \cfg{figure:atomic_pairhopping} show the spin flip
term and the pair hopping susceptibility in the atomic limit. Again, we observe that worm sampling is able to reproduce the analytic expression.

\begin{figure}
\centering
\includegraphics[scale=0.6]{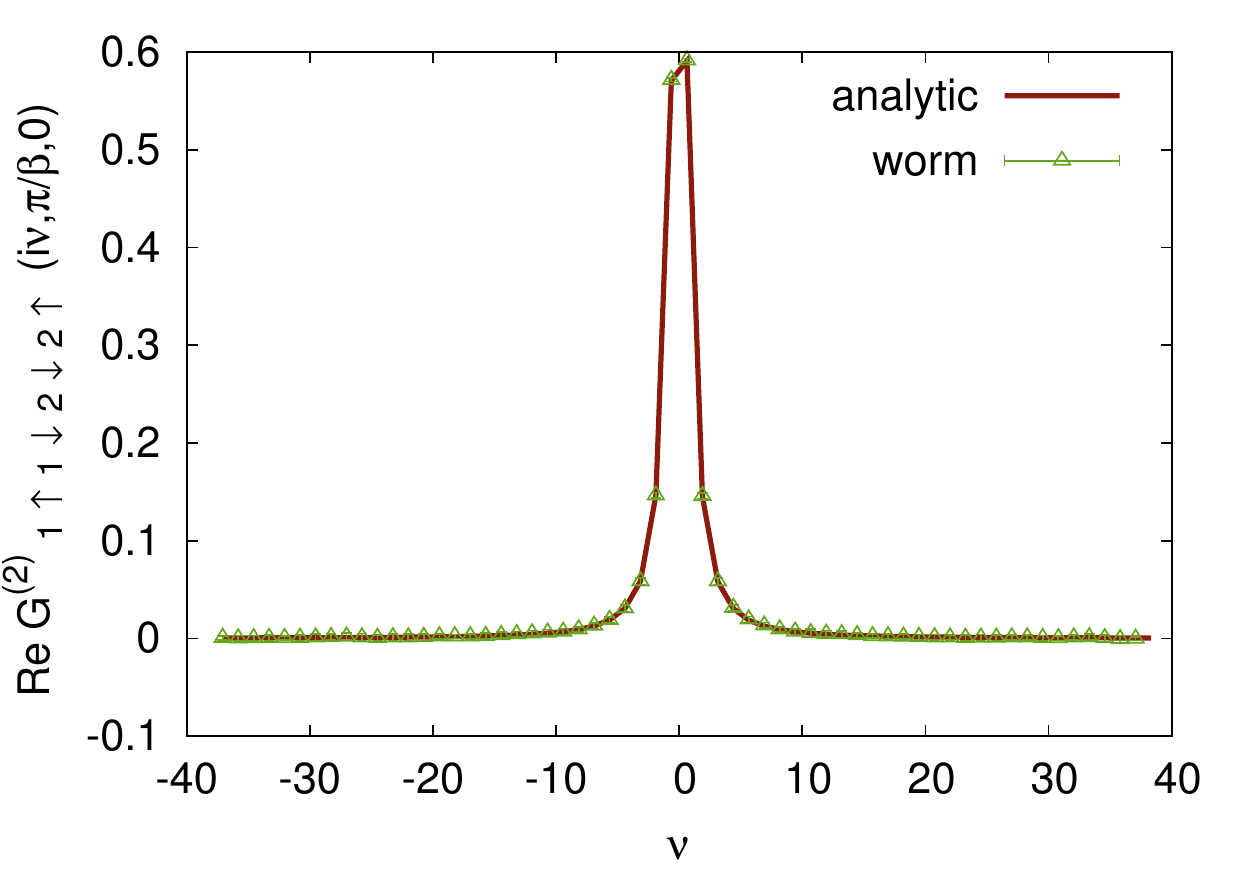}
\caption{Two-particle  spin flip Green's function $G^{(2)}_{1 \uparrow 1\downarrow 2\downarrow 2\uparrow}(i \nu, \pi/\beta, 0)$ vs.\  $i\nu$ for the atomic limit of the two-orbital AIM at 
$\beta = 5/D$, $U=1.0D$, $J=0.4D$, $U'=0.2D$, and $\mu=0.5D$ (half-filling). The balancing parameter was set to $\eta^{(2)}=0.09$. 
In the absence of any hybridization function, the worm algorithm (green triangles) is able to reproduce the analytic result (red line).} 
\label{figure:atomic_spinflip}
\end{figure}

\begin{figure}
\centering
\includegraphics[scale=0.6]{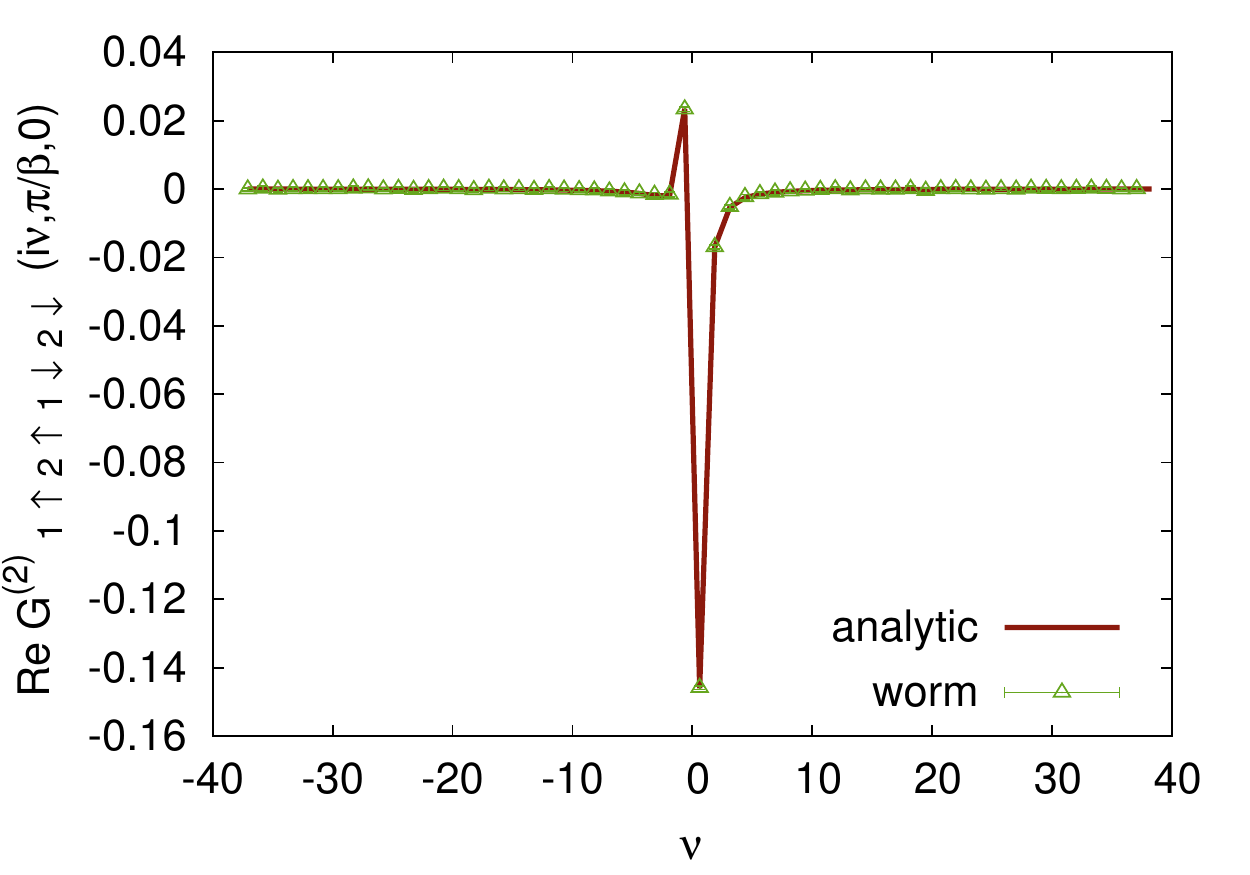}
\caption{Same as Fig. \ref{figure:atomic_spinflip} but for the
 pair hopping Green's function $G^{(2)}_{1 \uparrow 2\uparrow 1\downarrow 2\downarrow}(i \nu, \pi/\beta, 0)$.} 
\label{figure:atomic_pairhopping}
\end{figure}

So far we have only presented results for the spin flip and pair hopping term using worm sampling in the absence of a hybridization. While this atomic limit is very useful for benchmarking purposes, we are ultimately
interested in intermediate parameters, where CT-QMC algorithms are predominantly used, especially for calculating multi-orbital systems. In order to further verify our results, we exploit the SU(2) symmetry of Slater-Kanamori-like
interaction, where $\langle S_z(\tau) S_z(0) \rangle = \langle S_x(\tau) S_x(0) \rangle$ holds. 

Using partition function sampling, we can calculate the spin susceptibility in z-direction in a straight-forward manner.
Note  that we can express  $S_z(\tau)=n^i_\uparrow (\tau) - n^i_\downarrow (\tau)$
in terms of density operators so that $\langle S_z(\tau) S_z(0) \rangle$
can eventually be sampled by removing  diagonal hybridization functions in
partition function sampling.

This is not possible  for $\langle S_x(\tau) S_x(0) \rangle$ which is  expressed in terms of  spin flip two-particle Green's functions. While this 
 cannot be calculated in conventional partition function sampling, we can do so by using worm sampling. Instead of looking at the imaginary-time resolved
spin susceptibility, we verify the SU(2)-symmetry for the local spin susceptibility in terms of its Fourier transform to Matsubara frequencies $\chi_{\rm loc}(i \omega) = \int_0^\beta {\rm d} \tau e^{-i\omega \tau} \langle S_{z(x)}(\tau) S_{z(x)}(0) \rangle$.

\cfg{figure:susc_tot} shows the
spin-susceptibilities for the two-orbital AIM on a Bethe lattice. The worm sampling estimate for the  $ S_x S_x$ susceptibility in $x$-direction agrees with the  $ S_z S_z$ susceptibility in $z$-direction, 
which can be calculated both by worm and partition function sampling.
This further demonstrates the power of worm sampling to calculate general Green's functions and susceptibilities.

\begin{figure}
\centering
\includegraphics[scale=0.6]{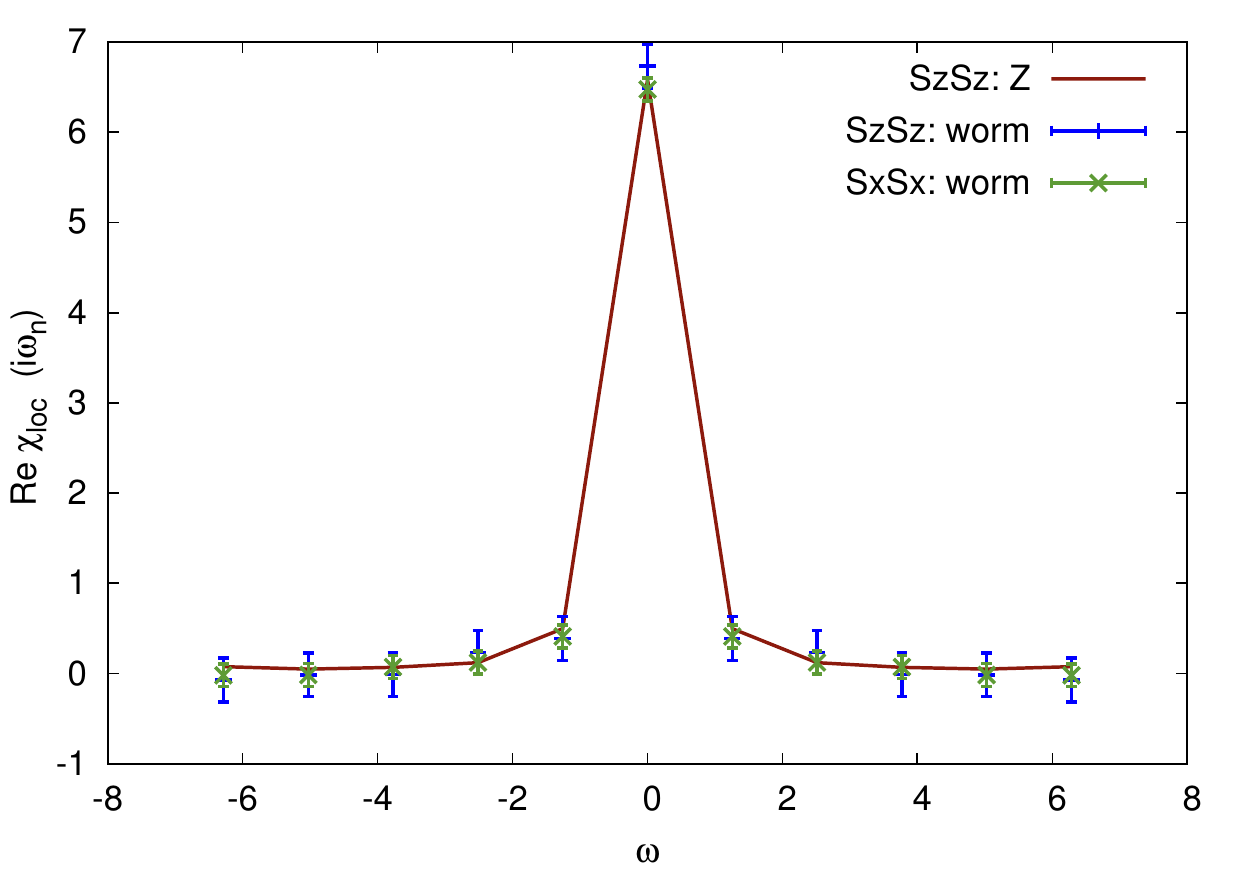}
\caption{Local spin susceptibility ${\rm Re} \chi_{loc}(i \omega)$ of the two-orbital AIM as a function of the bosonic Matsubara frequency $i\omega$. Parameters: identical semi-elliptic bands of
half-bandwidth $D$, 
$\beta = 5/D$,  $U=1.0D$, $J=0.4D$, $U'=0.2D$, and $\mu=0.5D$ (half-filling). The balancing parameter was set to $\eta^{(2)}=0.08$.
The SU(2) symmetry is conserved, as the $S_x S_x$ susceptibility  of the worm algorithm (green error bars) agrees well with the $S_z S_z$  susceptibility of partition function sampling (red line) and worm algorithm (blue error bars).} 
\label{figure:susc_tot}
\end{figure}

\section{Conclusion}\label{sec:conclusion}
In this work we have demonstrated how worm sampling provides a solution to some systematic failures of conventional CT-HYB algorithms. By inserting operators explicitly into the local trace,
we decouple the Green's function measurement from the hybridization function. This allows us to measure the one-particle and the two-particle Green's functions in situations, where the hybridization function is vanishing.
Further, we are able to generate off-diagonal components of the two-particle Green's function (spin flip and pair hopping terms). We have verified the algorithm by testing the atomic limit and showing
the SU(2) symmetry for a two-orbital Bethe model. The worm algorithm supplements the hybridization expansion CT-QMC solver with a numerically exact procedure for estimating two-particle correlation functions.

\acknowledgments
We thank G. Rohringer and P. Thunstr\"om for valuable discussions.
This work has been supported  by the  Vienna Scientific Cluster (VSC) Research Center funded by the Austrian Federal Ministry of Science, Research and Economy (bmwfw), DFG research unit FOR 1346, and the European Research Council under the European Union's Seventh Framework Programme (FP/2007-2013)/ERC through grant agreement n. 306447 (AbinitioD$\Gamma$A).  
A. H. and G.S. have been supported by the DFG (through SFB 1170 ``ToCoTronics'').
 The computational results presented have been achieved using the VSC.

\bibliography{bibliography}

\begin{thebibliography}{41}
\expandafter\ifx\csname natexlab\endcsname\relax\def\natexlab#1{#1}\fi
\expandafter\ifx\csname bibnamefont\endcsname\relax
  \def\bibnamefont#1{#1}\fi
\expandafter\ifx\csname bibfnamefont\endcsname\relax
  \def\bibfnamefont#1{#1}\fi
\expandafter\ifx\csname citenamefont\endcsname\relax
  \def\citenamefont#1{#1}\fi
\expandafter\ifx\csname url\endcsname\relax
  \def\url#1{\texttt{#1}}\fi
\expandafter\ifx\csname urlprefix\endcsname\relax\def\urlprefix{URL }\fi
\providecommand{\bibinfo}[2]{#2}
\providecommand{\eprint}[2][]{\url{#2}}

\bibitem[{\citenamefont{Anderson}(1961)}]{Anderson}
\bibinfo{author}{\bibfnamefont{P.~W.} \bibnamefont{Anderson}},
  \bibinfo{journal}{Phys. Rev.} \textbf{\bibinfo{volume}{124}},
  \bibinfo{pages}{41} (\bibinfo{year}{1961}),
  \urlprefix\url{http://link.aps.org/doi/10.1103/PhysRev.124.41}.

\bibitem[{\citenamefont{Hewson}(1997)}]{Hewson}
\bibinfo{author}{\bibfnamefont{A.~C.} \bibnamefont{Hewson}},
  \emph{\bibinfo{title}{The Kondo Problem to Heavy Fermions (Cambridge Studies
  in Magnetism)}} (\bibinfo{publisher}{Cambridge University Press},
  \bibinfo{year}{1997}), ISBN \bibinfo{isbn}{0521599474}.

\bibitem[{\citenamefont{Glazman and E.}(1988)}]{Glazman}
\bibinfo{author}{\bibfnamefont{L.~I.} \bibnamefont{Glazman}} \bibnamefont{and}
  \bibinfo{author}{\bibfnamefont{R.~M.} \bibnamefont{E.}},
  \bibinfo{journal}{JETP Lett.} \textbf{\bibinfo{volume}{47}},
  \bibinfo{pages}{452} (\bibinfo{year}{1988}),
  \urlprefix\url{http://www.jetpletters.ac.ru/ps/1095/article_16538.shtml}.

\bibitem[{\citenamefont{Ng and Lee}(1988)}]{Ng}
\bibinfo{author}{\bibfnamefont{T.~K.} \bibnamefont{Ng}} \bibnamefont{and}
  \bibinfo{author}{\bibfnamefont{P.~A.} \bibnamefont{Lee}},
  \bibinfo{journal}{Phys. Rev. Lett.} \textbf{\bibinfo{volume}{61}},
  \bibinfo{pages}{1768} (\bibinfo{year}{1988}),
  \urlprefix\url{http://link.aps.org/doi/10.1103/PhysRevLett.61.1768}.

\bibitem[{\citenamefont{Kouwenhoven and Glazman}(2001)}]{Kouwenhoven0}
\bibinfo{author}{\bibfnamefont{L.}~\bibnamefont{Kouwenhoven}} \bibnamefont{and}
  \bibinfo{author}{\bibfnamefont{L.}~\bibnamefont{Glazman}},
  \bibinfo{journal}{Physics World}  (\bibinfo{year}{2001}).

\bibitem[{\citenamefont{Madhavan et~al.}(1998)\citenamefont{Madhavan, Chen,
  Jamneala, Crommie, and Wingreen}}]{Madhavan}
\bibinfo{author}{\bibfnamefont{V.}~\bibnamefont{Madhavan}},
  \bibinfo{author}{\bibfnamefont{W.}~\bibnamefont{Chen}},
  \bibinfo{author}{\bibfnamefont{T.}~\bibnamefont{Jamneala}},
  \bibinfo{author}{\bibfnamefont{M.~F.} \bibnamefont{Crommie}},
  \bibnamefont{and} \bibinfo{author}{\bibfnamefont{N.~S.}
  \bibnamefont{Wingreen}}, \bibinfo{journal}{Science}
  \textbf{\bibinfo{volume}{280}}, \bibinfo{pages}{567} (\bibinfo{year}{1998}),
  \eprint{http://www.sciencemag.org/content/280/5363/567.full.pdf},
  \urlprefix\url{http://www.sciencemag.org/content/280/5363/567.abstract}.

\bibitem[{\citenamefont{Li et~al.}(1998)\citenamefont{Li, Schneider, Berndt,
  and Delley}}]{Li}
\bibinfo{author}{\bibfnamefont{J.}~\bibnamefont{Li}},
  \bibinfo{author}{\bibfnamefont{W.}~\bibnamefont{Schneider}},
  \bibinfo{author}{\bibfnamefont{R.}~\bibnamefont{Berndt}}, \bibnamefont{and}
  \bibinfo{author}{\bibfnamefont{B.}~\bibnamefont{Delley}},
  \bibinfo{journal}{Phys. Rev. Lett.} \textbf{\bibinfo{volume}{80}},
  \bibinfo{pages}{2893} (\bibinfo{year}{1998}),
  \urlprefix\url{http://link.aps.org/doi/10.1103/PhysRevLett.80.2893}.

\bibitem[{\citenamefont{Metzner and Vollhardt}(1989)}]{Metzner}
\bibinfo{author}{\bibfnamefont{W.}~\bibnamefont{Metzner}} \bibnamefont{and}
  \bibinfo{author}{\bibfnamefont{D.}~\bibnamefont{Vollhardt}},
  \bibinfo{journal}{Phys. Rev. Lett.} \textbf{\bibinfo{volume}{62}},
  \bibinfo{pages}{324} (\bibinfo{year}{1989}),
  \urlprefix\url{http://link.aps.org/doi/10.1103/PhysRevLett.62.324}.

\bibitem[{\citenamefont{Georges et~al.}(1996)\citenamefont{Georges, Kotliar,
  Krauth, and Rozenberg}}]{Georges}
\bibinfo{author}{\bibfnamefont{A.}~\bibnamefont{Georges}},
  \bibinfo{author}{\bibfnamefont{G.}~\bibnamefont{Kotliar}},
  \bibinfo{author}{\bibfnamefont{W.}~\bibnamefont{Krauth}}, \bibnamefont{and}
  \bibinfo{author}{\bibfnamefont{M.~J.} \bibnamefont{Rozenberg}},
  \bibinfo{journal}{Rev. Mod. Phys.} \textbf{\bibinfo{volume}{68}},
  \bibinfo{pages}{13} (\bibinfo{year}{1996}),
  \urlprefix\url{http://link.aps.org/doi/10.1103/RevModPhys.68.13}.

\bibitem[{\citenamefont{Kotliar and Vollhardt}(2004)}]{Kotliar_dmft}
\bibinfo{author}{\bibfnamefont{G.}~\bibnamefont{Kotliar}} \bibnamefont{and}
  \bibinfo{author}{\bibfnamefont{D.}~\bibnamefont{Vollhardt}},
  \bibinfo{journal}{Physics Today} \textbf{\bibinfo{volume}{57}},
  \bibinfo{pages}{53} (\bibinfo{year}{2004}),
  \urlprefix\url{http://dx.doi.org/10.1063/1.1712502}.

\bibitem[{\citenamefont{Held}(2007)}]{Held}
\bibinfo{author}{\bibfnamefont{K.}~\bibnamefont{Held}},
  \bibinfo{journal}{Advances in Physics} \textbf{\bibinfo{volume}{56}},
  \bibinfo{pages}{829} (\bibinfo{year}{2007}),
  \eprint{http://dx.doi.org/10.1080/00018730701619647},
  \urlprefix\url{http://dx.doi.org/10.1080/00018730701619647}.

\bibitem[{\citenamefont{Rubtsov and Lichtenstein}(2004)}]{Rubstov_ct_int}
\bibinfo{author}{\bibfnamefont{A.}~\bibnamefont{Rubtsov}} \bibnamefont{and}
  \bibinfo{author}{\bibfnamefont{A.}~\bibnamefont{Lichtenstein}},
  \bibinfo{journal}{Journal of Experimental and Theoretical Physics Letters}
  \textbf{\bibinfo{volume}{80}}, \bibinfo{pages}{61} (\bibinfo{year}{2004}),
  ISSN \bibinfo{issn}{0021-3640},
  \urlprefix\url{http://dx.doi.org/10.1134/1.1800216}.

\bibitem[{\citenamefont{Werner et~al.}(2006)\citenamefont{Werner, Comanac, de'
  Medici, Troyer, and Millis}}]{Werner_qmc}
\bibinfo{author}{\bibfnamefont{P.}~\bibnamefont{Werner}},
  \bibinfo{author}{\bibfnamefont{A.}~\bibnamefont{Comanac}},
  \bibinfo{author}{\bibfnamefont{L.}~\bibnamefont{de' Medici}},
  \bibinfo{author}{\bibfnamefont{M.}~\bibnamefont{Troyer}}, \bibnamefont{and}
  \bibinfo{author}{\bibfnamefont{A.~J.} \bibnamefont{Millis}},
  \bibinfo{journal}{Phys. Rev. Lett.} \textbf{\bibinfo{volume}{97}},
  \bibinfo{pages}{076405} (\bibinfo{year}{2006}),
  \urlprefix\url{http://link.aps.org/doi/10.1103/PhysRevLett.97.076405}.

\bibitem[{\citenamefont{Werner and Millis}(2006)}]{Werner}
\bibinfo{author}{\bibfnamefont{P.}~\bibnamefont{Werner}} \bibnamefont{and}
  \bibinfo{author}{\bibfnamefont{A.~J.} \bibnamefont{Millis}},
  \bibinfo{journal}{Phys. Rev. B} \textbf{\bibinfo{volume}{74}},
  \bibinfo{pages}{155107} (\bibinfo{year}{2006}),
  \urlprefix\url{http://link.aps.org/doi/10.1103/PhysRevB.74.155107}.

\bibitem[{\citenamefont{Gull et~al.}(2008)\citenamefont{Gull, Werner,
  Parcollet, and Troyer}}]{Gull_aux}
\bibinfo{author}{\bibfnamefont{E.}~\bibnamefont{Gull}},
  \bibinfo{author}{\bibfnamefont{P.}~\bibnamefont{Werner}},
  \bibinfo{author}{\bibfnamefont{O.}~\bibnamefont{Parcollet}},
  \bibnamefont{and} \bibinfo{author}{\bibfnamefont{M.}~\bibnamefont{Troyer}},
  \bibinfo{journal}{EPL (Europhysics Letters)} \textbf{\bibinfo{volume}{82}},
  \bibinfo{pages}{57003} (\bibinfo{year}{2008}),
  \urlprefix\url{http://stacks.iop.org/0295-5075/82/i=5/a=57003}.

\bibitem[{\citenamefont{Gull et~al.}(2011)\citenamefont{Gull, Millis,
  Lichtenstein, Rubtsov, Troyer, and Werner}}]{Gull}
\bibinfo{author}{\bibfnamefont{E.}~\bibnamefont{Gull}},
  \bibinfo{author}{\bibfnamefont{A.~J.} \bibnamefont{Millis}},
  \bibinfo{author}{\bibfnamefont{A.~I.} \bibnamefont{Lichtenstein}},
  \bibinfo{author}{\bibfnamefont{A.~N.} \bibnamefont{Rubtsov}},
  \bibinfo{author}{\bibfnamefont{M.}~\bibnamefont{Troyer}}, \bibnamefont{and}
  \bibinfo{author}{\bibfnamefont{P.}~\bibnamefont{Werner}},
  \bibinfo{journal}{Reviews of Modern Physics} \textbf{\bibinfo{volume}{83}},
  \bibinfo{pages}{349} (\bibinfo{year}{2011}),
  \urlprefix\url{http://dx.doi.org/10.1103/RevModPhys.83.349}.

\bibitem[{\citenamefont{Prokof’ev et~al.}(1996)\citenamefont{Prokof’ev,
  Svistunov, and Tupitsyn}}]{Prokofev_ct_orig}
\bibinfo{author}{\bibfnamefont{N.}~\bibnamefont{Prokof’ev}},
  \bibinfo{author}{\bibfnamefont{B.}~\bibnamefont{Svistunov}},
  \bibnamefont{and} \bibinfo{author}{\bibfnamefont{I.}~\bibnamefont{Tupitsyn}},
  \bibinfo{journal}{Journal of Experimental and Theoretical Physics Letters}
  \textbf{\bibinfo{volume}{64}}, \bibinfo{pages}{911} (\bibinfo{year}{1996}),
  ISSN \bibinfo{issn}{0021-3640},
  \urlprefix\url{http://dx.doi.org/10.1134/1.567243}.

\bibitem[{\citenamefont{Prokof'ev
  et~al.}(1998{\natexlab{a}})\citenamefont{Prokof'ev, Svistunov, and
  Tupitsyn}}]{Prokofev_ct}
\bibinfo{author}{\bibfnamefont{N.~V.} \bibnamefont{Prokof'ev}},
  \bibinfo{author}{\bibfnamefont{B.~V.} \bibnamefont{Svistunov}},
  \bibnamefont{and} \bibinfo{author}{\bibfnamefont{I.~S.}
  \bibnamefont{Tupitsyn}}, \bibinfo{journal}{Journal of Experimental and
  Theoretical Physics} \textbf{\bibinfo{volume}{87}}, \bibinfo{pages}{310}
  (\bibinfo{year}{1998}{\natexlab{a}}), ISSN \bibinfo{issn}{1063-7761},
  \urlprefix\url{http://dx.doi.org/10.1134/1.558661}.

\bibitem[{\citenamefont{Hirsch and Fye}(1986)}]{Hirsch}
\bibinfo{author}{\bibfnamefont{J.~E.} \bibnamefont{Hirsch}} \bibnamefont{and}
  \bibinfo{author}{\bibfnamefont{R.~M.} \bibnamefont{Fye}},
  \bibinfo{journal}{Phys. Rev. Lett.} \textbf{\bibinfo{volume}{56}},
  \bibinfo{pages}{2521} (\bibinfo{year}{1986}),
  \urlprefix\url{http://link.aps.org/doi/10.1103/PhysRevLett.56.2521}.

\bibitem[{\citenamefont{Sakai et~al.}(2006)\citenamefont{Sakai, Arita, Held,
  and Aoki}}]{Sakai}
\bibinfo{author}{\bibfnamefont{S.}~\bibnamefont{Sakai}},
  \bibinfo{author}{\bibfnamefont{R.}~\bibnamefont{Arita}},
  \bibinfo{author}{\bibfnamefont{K.}~\bibnamefont{Held}}, \bibnamefont{and}
  \bibinfo{author}{\bibfnamefont{H.}~\bibnamefont{Aoki}},
  \bibinfo{journal}{Phys. Rev. B} \textbf{\bibinfo{volume}{74}},
  \bibinfo{pages}{155102} (\bibinfo{year}{2006}),
  \urlprefix\url{http://link.aps.org/doi/10.1103/PhysRevB.74.155102}.

\bibitem[{\citenamefont{Bulla et~al.}(2008)\citenamefont{Bulla, Costi, and
  Pruschke}}]{Bulla}
\bibinfo{author}{\bibfnamefont{R.}~\bibnamefont{Bulla}},
  \bibinfo{author}{\bibfnamefont{T.~A.} \bibnamefont{Costi}}, \bibnamefont{and}
  \bibinfo{author}{\bibfnamefont{T.}~\bibnamefont{Pruschke}},
  \bibinfo{journal}{Rev. Mod. Phys.} \textbf{\bibinfo{volume}{80}},
  \bibinfo{pages}{395} (\bibinfo{year}{2008}),
  \urlprefix\url{http://link.aps.org/doi/10.1103/RevModPhys.80.395}.

\bibitem[{\citenamefont{Karski et~al.}(2008)\citenamefont{Karski, Raas, and
  Uhrig}}]{Karski}
\bibinfo{author}{\bibfnamefont{M.}~\bibnamefont{Karski}},
  \bibinfo{author}{\bibfnamefont{C.}~\bibnamefont{Raas}}, \bibnamefont{and}
  \bibinfo{author}{\bibfnamefont{G.~S.} \bibnamefont{Uhrig}},
  \bibinfo{journal}{Phys. Rev. B} \textbf{\bibinfo{volume}{77}},
  \bibinfo{pages}{075116} (\bibinfo{year}{2008}),
  \urlprefix\url{http://link.aps.org/doi/10.1103/PhysRevB.77.075116}.

\bibitem[{\citenamefont{Ganahl et~al.}(2014)\citenamefont{Ganahl, Thunstr\"om,
  Verstraete, Held, and Evertz}}]{Ganahl}
\bibinfo{author}{\bibfnamefont{M.}~\bibnamefont{Ganahl}},
  \bibinfo{author}{\bibfnamefont{P.}~\bibnamefont{Thunstr\"om}},
  \bibinfo{author}{\bibfnamefont{F.}~\bibnamefont{Verstraete}},
  \bibinfo{author}{\bibfnamefont{K.}~\bibnamefont{Held}}, \bibnamefont{and}
  \bibinfo{author}{\bibfnamefont{H.~G.} \bibnamefont{Evertz}},
  \bibinfo{journal}{Phys. Rev. B} \textbf{\bibinfo{volume}{90}},
  \bibinfo{pages}{045144} (\bibinfo{year}{2014}),
  \urlprefix\url{http://link.aps.org/doi/10.1103/PhysRevB.90.045144}.

\bibitem[{\citenamefont{Caffarel and Krauth}(1994)}]{Caffarel}
\bibinfo{author}{\bibfnamefont{M.}~\bibnamefont{Caffarel}} \bibnamefont{and}
  \bibinfo{author}{\bibfnamefont{W.}~\bibnamefont{Krauth}},
  \bibinfo{journal}{Phys. Rev. Lett.} \textbf{\bibinfo{volume}{72}},
  \bibinfo{pages}{1545} (\bibinfo{year}{1994}),
  \urlprefix\url{http://link.aps.org/doi/10.1103/PhysRevLett.72.1545}.

\bibitem[{\citenamefont{Zgid et~al.}(2012)\citenamefont{Zgid, Gull, and
  Chan}}]{Zgid}
\bibinfo{author}{\bibfnamefont{D.}~\bibnamefont{Zgid}},
  \bibinfo{author}{\bibfnamefont{E.}~\bibnamefont{Gull}}, \bibnamefont{and}
  \bibinfo{author}{\bibfnamefont{G.~K.-L.} \bibnamefont{Chan}},
  \bibinfo{journal}{Phys. Rev. B} \textbf{\bibinfo{volume}{86}},
  \bibinfo{pages}{165128} (\bibinfo{year}{2012}),
  \urlprefix\url{http://link.aps.org/doi/10.1103/PhysRevB.86.165128}.

\bibitem[{\citenamefont{Prokof'ev
  et~al.}(1998{\natexlab{b}})\citenamefont{Prokof'ev, Svistunov, and
  Tupitsyn}}]{Prokofev_hfye}
\bibinfo{author}{\bibfnamefont{N.~V.} \bibnamefont{Prokof'ev}},
  \bibinfo{author}{\bibfnamefont{B.}~\bibnamefont{Svistunov}},
  \bibnamefont{and} \bibinfo{author}{\bibfnamefont{I.}~\bibnamefont{Tupitsyn}},
  \bibinfo{journal}{Physics Letters A} \textbf{\bibinfo{volume}{238}},
  \bibinfo{pages}{253} (\bibinfo{year}{1998}{\natexlab{b}}),
  \urlprefix\url{http://dx.doi.org/10.1016/S0375-9601(97)00957-2}.

\bibitem[{\citenamefont{Burovski et~al.}(2006)\citenamefont{Burovski,
  Prokof'ev, Svistunov, and Troyer}}]{Burovski}
\bibinfo{author}{\bibfnamefont{E.}~\bibnamefont{Burovski}},
  \bibinfo{author}{\bibfnamefont{N.}~\bibnamefont{Prokof'ev}},
  \bibinfo{author}{\bibfnamefont{B.}~\bibnamefont{Svistunov}},
  \bibnamefont{and} \bibinfo{author}{\bibfnamefont{M.}~\bibnamefont{Troyer}},
  \bibinfo{journal}{New Journal of Physics} \textbf{\bibinfo{volume}{8}},
  \bibinfo{pages}{153} (\bibinfo{year}{2006}),
  \urlprefix\url{http://stacks.iop.org/1367-2630/8/i=8/a=153}.

\bibitem[{\citenamefont{Toschi et~al.}(2007)\citenamefont{Toschi, Katanin, and
  Held}}]{Toschi}
\bibinfo{author}{\bibfnamefont{A.}~\bibnamefont{Toschi}},
  \bibinfo{author}{\bibfnamefont{A.~A.} \bibnamefont{Katanin}},
  \bibnamefont{and} \bibinfo{author}{\bibfnamefont{K.}~\bibnamefont{Held}},
  \bibinfo{journal}{Phys. Rev. B} \textbf{\bibinfo{volume}{75}},
  \bibinfo{pages}{045118} (\bibinfo{year}{2007}),
  \urlprefix\url{http://link.aps.org/doi/10.1103/PhysRevB.75.045118}.

\bibitem[{\citenamefont{Rubtsov et~al.}(2008)\citenamefont{Rubtsov, Katsnelson,
  and Lichtenstein}}]{Rubtsov}
\bibinfo{author}{\bibfnamefont{A.~N.} \bibnamefont{Rubtsov}},
  \bibinfo{author}{\bibfnamefont{M.~I.} \bibnamefont{Katsnelson}},
  \bibnamefont{and} \bibinfo{author}{\bibfnamefont{A.~I.}
  \bibnamefont{Lichtenstein}}, \bibinfo{journal}{Phys. Rev. B}
  \textbf{\bibinfo{volume}{77}}, \bibinfo{pages}{033101}
  (\bibinfo{year}{2008}),
  \urlprefix\url{http://link.aps.org/doi/10.1103/PhysRevB.77.033101}.

\bibitem[{\citenamefont{Rohringer et~al.}(2013)\citenamefont{Rohringer, Toschi,
  Hafermann, Held, Anisimov, and Katanin}}]{Rohringer_1PI}
\bibinfo{author}{\bibfnamefont{G.}~\bibnamefont{Rohringer}},
  \bibinfo{author}{\bibfnamefont{A.}~\bibnamefont{Toschi}},
  \bibinfo{author}{\bibfnamefont{H.}~\bibnamefont{Hafermann}},
  \bibinfo{author}{\bibfnamefont{K.}~\bibnamefont{Held}},
  \bibinfo{author}{\bibfnamefont{V.~I.} \bibnamefont{Anisimov}},
  \bibnamefont{and} \bibinfo{author}{\bibfnamefont{A.~A.}
  \bibnamefont{Katanin}}, \bibinfo{journal}{Phys. Rev. B}
  \textbf{\bibinfo{volume}{88}}, \bibinfo{pages}{115112}
  (\bibinfo{year}{2013}),
  \urlprefix\url{http://link.aps.org/doi/10.1103/PhysRevB.88.115112}.

\bibitem[{\citenamefont{Taranto et~al.}(2014)\citenamefont{Taranto,
  Andergassen, Bauer, Held, Katanin, Metzner, Rohringer, and Toschi}}]{Taranto}
\bibinfo{author}{\bibfnamefont{C.}~\bibnamefont{Taranto}},
  \bibinfo{author}{\bibfnamefont{S.}~\bibnamefont{Andergassen}},
  \bibinfo{author}{\bibfnamefont{J.}~\bibnamefont{Bauer}},
  \bibinfo{author}{\bibfnamefont{K.}~\bibnamefont{Held}},
  \bibinfo{author}{\bibfnamefont{A.}~\bibnamefont{Katanin}},
  \bibinfo{author}{\bibfnamefont{W.}~\bibnamefont{Metzner}},
  \bibinfo{author}{\bibfnamefont{G.}~\bibnamefont{Rohringer}},
  \bibnamefont{and} \bibinfo{author}{\bibfnamefont{A.}~\bibnamefont{Toschi}},
  \bibinfo{journal}{Phys. Rev. Lett.} \textbf{\bibinfo{volume}{112}},
  \bibinfo{pages}{196402} (\bibinfo{year}{2014}),
  \urlprefix\url{http://link.aps.org/doi/10.1103/PhysRevLett.112.196402}.

\bibitem[{\citenamefont{Hafermann et~al.}(2012)\citenamefont{Hafermann, Patton,
  and Werner}}]{Hafermann}
\bibinfo{author}{\bibfnamefont{H.}~\bibnamefont{Hafermann}},
  \bibinfo{author}{\bibfnamefont{K.~R.} \bibnamefont{Patton}},
  \bibnamefont{and} \bibinfo{author}{\bibfnamefont{P.}~\bibnamefont{Werner}},
  \bibinfo{journal}{Phys. Rev. B} \textbf{\bibinfo{volume}{85}},
  \bibinfo{pages}{205106} (\bibinfo{year}{2012}),
  \urlprefix\url{http://link.aps.org/doi/10.1103/PhysRevB.85.205106}.

\bibitem[{\citenamefont{Boehnke et~al.}(2011)\citenamefont{Boehnke, Hafermann,
  Ferrero, Lechermann, and Parcollet}}]{Boehnke}
\bibinfo{author}{\bibfnamefont{L.}~\bibnamefont{Boehnke}},
  \bibinfo{author}{\bibfnamefont{H.}~\bibnamefont{Hafermann}},
  \bibinfo{author}{\bibfnamefont{M.}~\bibnamefont{Ferrero}},
  \bibinfo{author}{\bibfnamefont{F.}~\bibnamefont{Lechermann}},
  \bibnamefont{and}
  \bibinfo{author}{\bibfnamefont{O.}~\bibnamefont{Parcollet}},
  \bibinfo{journal}{Phys. Rev. B} \textbf{\bibinfo{volume}{84}},
  \bibinfo{pages}{075145} (\bibinfo{year}{2011}),
  \urlprefix\url{http://link.aps.org/doi/10.1103/PhysRevB.84.075145}.

\bibitem[{\citenamefont{Augustinsky and Kunes}(2013)}]{Augustinsky}
\bibinfo{author}{\bibfnamefont{P.}~\bibnamefont{Augustinsky}} \bibnamefont{and}
  \bibinfo{author}{\bibfnamefont{J.}~\bibnamefont{Kunes}},
  \bibinfo{journal}{Computer Physics Communications}
  \textbf{\bibinfo{volume}{184}}, \bibinfo{pages}{2119 }
  (\bibinfo{year}{2013}), ISSN \bibinfo{issn}{0010-4655},
  \urlprefix\url{http://www.sciencedirect.com/science/article/pii/S0010465513001409}.

\bibitem[{\citenamefont{Seth et~al.}()\citenamefont{Seth, Krivenko, Ferrero,
  and Parcollet}}]{Olivier}
\bibinfo{author}{\bibfnamefont{P.}~\bibnamefont{Seth}},
  \bibinfo{author}{\bibfnamefont{I.}~\bibnamefont{Krivenko}},
  \bibinfo{author}{\bibfnamefont{M.}~\bibnamefont{Ferrero}}, \bibnamefont{and}
  \bibinfo{author}{\bibfnamefont{O.}~\bibnamefont{Parcollet}},
  \eprint{1507.00175}, \urlprefix\url{http://arxiv.org/abs/1507.00175}.

\bibitem[{\citenamefont{Gull et~al.}(2010)\citenamefont{Gull, Reichman, and
  Millis}}]{Gull_bold}
\bibinfo{author}{\bibfnamefont{E.}~\bibnamefont{Gull}},
  \bibinfo{author}{\bibfnamefont{D.~R.} \bibnamefont{Reichman}},
  \bibnamefont{and} \bibinfo{author}{\bibfnamefont{A.~J.}
  \bibnamefont{Millis}}, \bibinfo{journal}{Phys. Rev. B}
  \textbf{\bibinfo{volume}{82}}, \bibinfo{pages}{075109}
  (\bibinfo{year}{2010}),
  \urlprefix\url{http://link.aps.org/doi/10.1103/PhysRevB.82.075109}.

\bibitem[{\citenamefont{Rohringer et~al.}(2012)\citenamefont{Rohringer, Valli,
  and Toschi}}]{Rohringer}
\bibinfo{author}{\bibfnamefont{G.}~\bibnamefont{Rohringer}},
  \bibinfo{author}{\bibfnamefont{A.}~\bibnamefont{Valli}}, \bibnamefont{and}
  \bibinfo{author}{\bibfnamefont{A.}~\bibnamefont{Toschi}},
  \bibinfo{journal}{Phys. Rev. B} \textbf{\bibinfo{volume}{86}},
  \bibinfo{pages}{125114} (\bibinfo{year}{2012}),
  \urlprefix\url{http://link.aps.org/doi/10.1103/PhysRevB.86.125114}.

\bibitem[{\citenamefont{Hafermann et~al.}(2009)\citenamefont{Hafermann, Jung,
  Brener, Katsnelson, Rubtsov, and Lichtenstein}}]{Hafermann_atomic}
\bibinfo{author}{\bibfnamefont{H.}~\bibnamefont{Hafermann}},
  \bibinfo{author}{\bibfnamefont{C.}~\bibnamefont{Jung}},
  \bibinfo{author}{\bibfnamefont{S.}~\bibnamefont{Brener}},
  \bibinfo{author}{\bibfnamefont{M.~I.} \bibnamefont{Katsnelson}},
  \bibinfo{author}{\bibfnamefont{A.~N.} \bibnamefont{Rubtsov}},
  \bibnamefont{and} \bibinfo{author}{\bibfnamefont{A.~I.}
  \bibnamefont{Lichtenstein}}, \bibinfo{journal}{EPL (Europhysics Letters)}
  \textbf{\bibinfo{volume}{85}}, \bibinfo{pages}{27007} (\bibinfo{year}{2009}),
  \urlprefix\url{http://stacks.iop.org/0295-5075/85/i=2/a=27007}.

\bibitem[{\citenamefont{Haule}(2007)}]{Haule_2007}
\bibinfo{author}{\bibfnamefont{K.}~\bibnamefont{Haule}},
  \bibinfo{journal}{Phys. Rev. B} \textbf{\bibinfo{volume}{75}},
  \bibinfo{pages}{155113} (\bibinfo{year}{2007}),
  \urlprefix\url{http://link.aps.org/doi/10.1103/PhysRevB.75.155113}.

\bibitem[{\citenamefont{Parragh et~al.}(2012)\citenamefont{Parragh, Toschi,
  Held, and Sangiovanni}}]{Parragh}
\bibinfo{author}{\bibfnamefont{N.}~\bibnamefont{Parragh}},
  \bibinfo{author}{\bibfnamefont{A.}~\bibnamefont{Toschi}},
  \bibinfo{author}{\bibfnamefont{K.}~\bibnamefont{Held}}, \bibnamefont{and}
  \bibinfo{author}{\bibfnamefont{G.}~\bibnamefont{Sangiovanni}},
  \bibinfo{journal}{Phys. Rev. B} \textbf{\bibinfo{volume}{86}},
  \bibinfo{pages}{155158} (\bibinfo{year}{2012}),
  \urlprefix\url{http://link.aps.org/doi/10.1103/PhysRevB.86.155158}.

\bibitem[{\citenamefont{Kanamori}(1963)}]{Kanamori}
\bibinfo{author}{\bibfnamefont{J.}~\bibnamefont{Kanamori}},
  \bibinfo{journal}{Progress of Theoretical Physics}
  \textbf{\bibinfo{volume}{30}}, \bibinfo{pages}{275} (\bibinfo{year}{1963}),
  \eprint{http://ptp.oxfordjournals.org/content/30/3/275},
  \urlprefix\url{http://ptp.oxfordjournals.org/content/30/3/275.abstract}.

\end{thebibliography}

\vfill\eject

\end{document}